\documentclass[aps,pra,twocolumn,floatfix,reprint,showpacs,superscriptaddress]{revtex4-1}
\usepackage{amsmath}
\usepackage{color}
\usepackage{graphicx,graphics}
\usepackage{braket}
\begin{document}

\title{Cross-entangling electronic and nuclear spins of distant nitrogen-vacancy centers in noisy environments by means of quantum microwave radiation}

\author{A. Viviana G\'omez}
\email{av.gomez176@uniandes.edu.co}
\homepage[]{Present address: Civil Engineer Department, Universidad Mariana, San Juan de Pasto, Colombia.}
\affiliation{Departamento de F\'{\i}sica, Universidad de Los Andes, A.A.4976, Bogot\'a D.C., Colombia}

\author{Ferney J. Rodr\'{\i}guez}
\affiliation{Departamento de F\'{\i}sica, Universidad de Los Andes, A.A.4976, Bogot\'a D.C., Colombia}

\author{Luis Quiroga}
\affiliation{Departamento de F\'{\i}sica, Universidad de Los Andes, A.A.4976, Bogot\'a D.C., Colombia}

\date{\today }
\begin{abstract}
Nitrogen-vacancy (NV) defect centers in diamond are strong candidates to generate entangled states in solid-state environments even at room temperature. Quantum correlations in spatially separated NV systems, for distances between NVs ranging from a few nanometers to a few kilometers, have been recently reported.
In the present work we consider the entanglement transfer from two-mode microwave squeezed (entangled) photons, which are in resonance with the two lowest NV electron spin states, to initially unentangled NV centers. We first demonstrate that the entanglement transfer process from quantum microwaves to isolated NV electron spins is feasible.
We then proceed to extend the previous results to more realistic scenarios where $^{13}$C nuclear spin baths surrounding each NV are included, quantifying the entanglement transfer efficiency and robustness under the effects of dephasing/dissipation noisy nuclear baths. Finally, we address the issue of assessing the possibility of entanglement transfer from the squeezed microwave light to two remote nuclear spins closely linked to different NV centers.
\end{abstract}

\maketitle

\section{INTRODUCTION}
Recently a great deal of interest has arisen in quantum systems operating in the microwave sector of the electromagnetic spectrum since they provide
new opportunities for exploring fundamental aspects of quantum physics as well as possible applications in the field of quantum information and computation. Important steps in profiting microwave active quantum architectures include superconducting (SC) circuits ~\cite{Burkard2004,Makhlin2001,Wendin2007,You2003,You2011} and the manipulation of nuclear and electronic spins in solids ~\cite{Morello2010,Guillot2006,Doherty2012,Wrachtrup2006,Simmons2011,Tyryshkin2011}. A promising idea
pursued by several groups is to combine different matter subsystems in a hybrid quantum system to take advantage of the scalability, flexibility and large coupling to microwave fields of some of them, for instance SC circuits, and to exploit large coherence times of other subsystems, such as solid-state spin systems, for storing quantum information in stable quantum registers ~\cite{Wach,Wallquist2009}. From this perspective, nuclear spins prove more suitable than electronic spins. However, the direct control of spatially distant nuclear spins is challenging due to the weak coupling between themselves. Thus, the search for nuclear long-range entangling mechanisms which allow for opportunities to overcome those limitations are of great interest.

An excellent platform for undertaking that search is provided by nitrogen-vacancy (NV) centers in diamond. A single NV center is a well characterized defect in diamond consisting of a substitutional nitrogen atom next to a carbon vacancy in an adjacent lattice site ~\cite{Popa2004}. It has been demonstrated the selective addressing and controlling of a single NV, even at room temperature, and how their constituent electronic and nuclear spins can be effectively manipulated and
potentially coupled together ~\cite{Childress20,Dobrovitski2009}. On the other hand, the dipolar and hyperfine interactions between the electronic and nuclear spins in NV centers have been extensively studied. Individual control and readout of nuclear spin qubits coupled to the electronic spin has been demonstrated ~\cite{Hanson2016}. Besides that, the control of two nuclear spins on an individual basis, generates entanglement of two $^{13}C$ nuclear spins at the first coordination shell of the vacancy ~\cite{Neumann2008} and  mediate the entanglement between multiple photons ~\cite{Rao2015}.

Numerous quantum information protocols with NV centers have been previously discussed in the literature. The quantum dynamics of distant $^{13}C$ nuclear spins has been probed using a weak coupling with the electronic spin in NV centers ~\cite{Kolkowitz12}. Furthermore, the initialization of electron and nuclear spin qubits \cite{Dutt2007}, the transfer of quantum states ~\cite{Dutt2007,Wrachtrup2016} and the generation of controlled quantum gate between distant nitrogen nuclear spins ~\cite{Auer2016} represent a step forward to build a quantum repeater network for long distances. An important issue in the field of quantum information is the generation of entangled states in a scalable way. The combination of radiation excitation from different wavelength sectors of the electromagnetic spectrum (optical, microwave and radio-frequency) has allowed to engineer protocols for reaching entanglement between electron spins in two separate NVs ~\cite{Neumann10}, the electron spin of a single NV and its neighbor nitrogen nucleus ~\cite{Cappellaro2009} or the NV electron and a closely placed $^{13}C$ nucleus ~\cite{Neumann2008}. Moreover,  other proposals show protocols to generate spin-photon entangled states between the ground state spin of a single NV center and the polarization of an emitted optical photon ~\cite{Togan2010}, heralded entanglement between solid-state qubits using optical photons ~\cite{Bernien13} and entanglement between NV electron spins separated up to $1.3$ Km have been reported ~\cite{Hensen15}.

\begin{figure}[t]
\begin{center}
\includegraphics[width=0.5\textwidth]{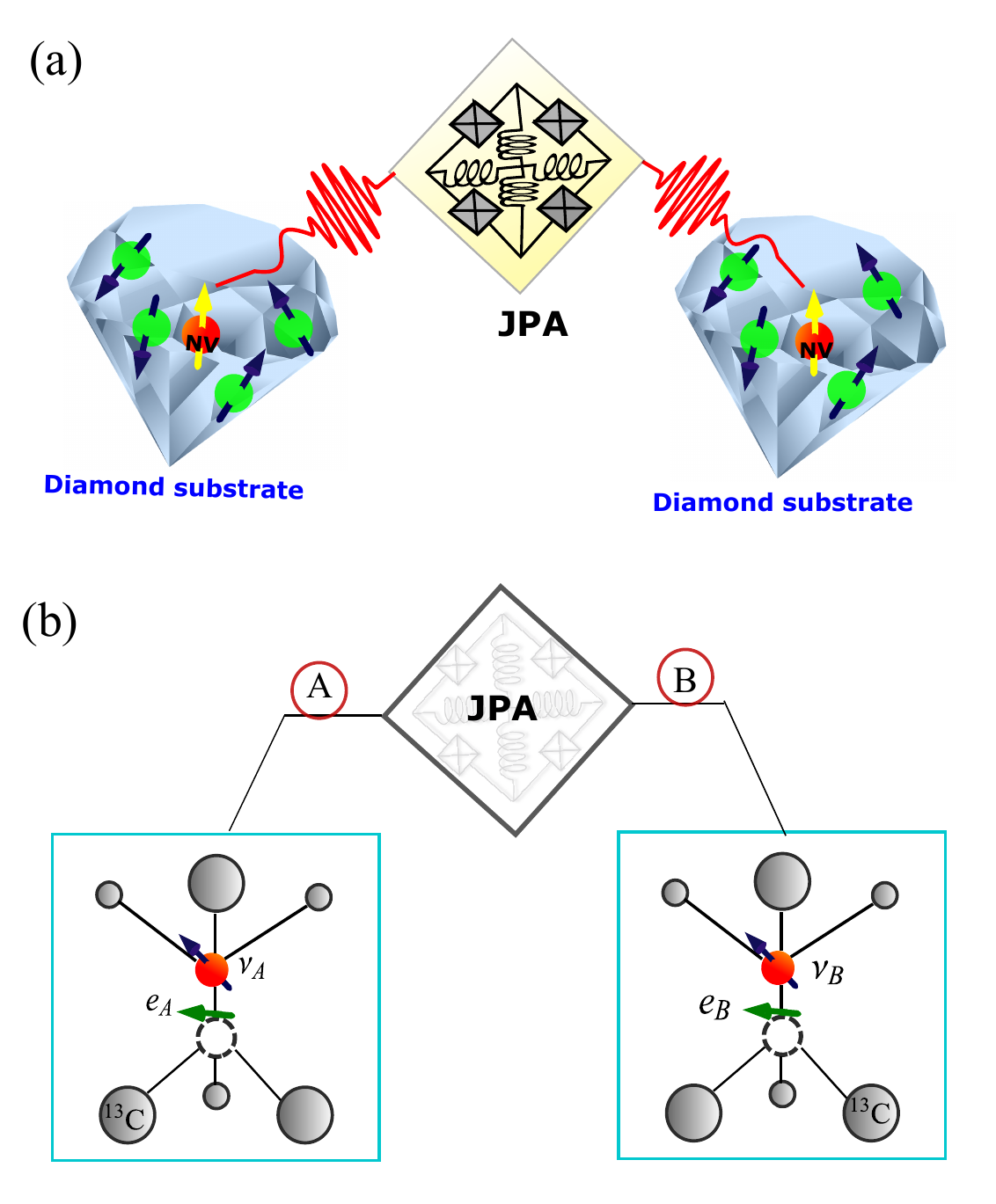}
\end{center}
\caption{a) Two distant single NV centers, each one embedded in its own nuclear spin bath, in different branches of a parametric Josephson amplifier producing highly entangled microwave photons. (b) Schematics of (a) where $e_{i}$, $\nu_{i}$ denote the electronic and nuclear spins of the individual NV center in branch $i$ ($i=A,B$).}
\label{fig:setup}
\end{figure}

In the present work we present a theoretical proposal based on NV defect centers in diamond to reach entanglement between distant electron and/or nuclear spins mediated by a quantum (squeezed) microwave field (QMF) as provided by a two-mode Josephson mixer ~\cite{Flurin12}, see Fig.~\ref{fig:setup}. The NV center has an electronic spin $S=1$ mostly localized at the defect bond. However, about 11\% of its electron spin density is distributed over the nearest neighbor carbon atoms and as a result substantial hyperfine and dipolar couplings with neighboring carbon nuclear spins ($^{13}C$) are sizeable \cite{Neumann2008}.
On the other hand, a diluted network of spin-$1/2$ $^{13}C$-nuclei forms a mesoscopic spin bath for a NV center. Under these conditions, we demonstrate that it is feasible the transfer of entanglement from the QMF to a pair of distant NVs (both electronic and nuclear spins) in such a noisy solid-state environment. First, we propose to entangle the electronic spins with a third party o mediator:  If the electronic spins are strongly coupled to their nearest nuclear spins, the hyperfine interaction between them allows an effective entanglement transfer to the nuclear spins.

Previous related works have proposed the use of NV centers as hybrid quantum systems ~\cite{Nori,Marcos,Zhu,Kubo} in which electron spins provide high fidelity control and readout while nuclear spins,
with ultra-long coherence times, support robust quantum registers. Also, the entanglement transfer from continuous variables to discrete spin systems has been considered from different approaches ~\cite{Gomez2016,Solano,Cirac,Paternostro04}. By contrast with most of previous studies, our present approach not only propose the entanglement generation between NV electronic spins but, most importantly, it also predicts the entanglement transfer to distant nuclear spins in noisy spin environments.

The paper is organized as follows: In Sect.\ \ref{sec:A} we address the entanglement transfer from a two-mode entangled QMF to the electronic NV-spins in noisy environments associated with nuclear spin baths. In Sect.\ \ref{sec:B} we extend previous results to the coupled electron-nuclear NV-spins by discussing three different scenarios:  two distant NV electron spins, two nuclear spins and one non-local electron-nucleus spin pair. A relevant result of this analysis is the identification of regimes for which  maximum entanglement is obtained in noisy environments. In Sect.\ \ref{sec:C}, we report numerical results for the time dependent entanglement generation and the identification of optimal parameters for maximum entanglement transfer under nuclear spin bath effects. Finally, in Sect.\ \ref{sec:Conclusions} we draw our conclusions and discuss some possible outlooks.
\section{QMF power entangling over two distant electron-electron spins in noisy NVs}
\label{sec:A}

The physics contained in the full system displayed in Fig.~\ref{fig:setup} is quite rich
and it is therefore instructive to consider a limiting case before
analyzing the full cross-entangling processes in the composite multi-bath environment. In the following, we derive and discuss separately results for the uncoupled electron-nucleus NV system, for short $e_{i}$-$\nu_{i}$ system, $i=A,B$, because of its high relevance for the existing theoretical and experimental literature.Thus, we start by considering the simplest scenario where we disregard the effects of the closest nuclear spin (see Fig.~\ref{fig:setup}-(b)): a two-arm device where in each path, $A$ and $B$, we place a single NV-electron driven by an entangled QMF in presence of a diluted $^{13}C$ nuclear noisy bath. In each path a microwave cavity enhances the NV-microwave field coupling strength. The subsystems labeled by $A$ and $B$ are assumed to be identical. We assume that a magnetic field is applied along the $z$ axis, leading to a Zeeman splitting between the electronic sub-levels with spin z-component $m_{s}=\pm 1$ ~\cite{Hanson2006}. In this way, the QMF should be quasi-resonant with the single $m_{s}=0$,$m_{s}=-1$ transition which will be described as an effective $1/2$-spin.

Although electronic and nuclear spins are well known for their long coherence times, for NV centers in diamond a major decoherence source is generated by the coupling between the central spin and other spins in the sample, such as electronic nitrogen spins or nuclear carbon spins. Here, we explore the influence of a $^{13}C$ spin bath on the entanglement transfer process.

Thus, for the uncoupled $e_{i}-\nu_{i}$ system the two-arm whole Hamiltonian is
\begin{eqnarray}
\nonumber \hat{H} &=&\sum_{j=A,B}\left[ \frac{\omega _{j}}{2}\hat{\sigma}
_{z,j}+\Omega _{j}\hat{a}_{j}^{\dag }\hat{a}_{j}+g_{j}\left( \hat{a}
_{j}^{\dag }\hat{\sigma}_{j}^{-}+\hat{a}_{j}\hat{\sigma}_{j}^{+}\right)
\right.\\
&&+\left. \hat{H}_{EB,j}+\hat{H}_{B,j} \right].
\label{Eq:m1}
\end{eqnarray}

The first three terms in Eq.(\ref{Eq:m1}) correspond to the usual Jaynes-Cummings (JC) Hamiltonian, where $\omega _{j}$ and $\Omega _{j}$ denote the electronic spin splitting and microwave cavity frequency, respectively and $g_{j}$ describe the electron-cavity coupling in arm $j$. The $\hat{\sigma}_{j,z}$ operator represents the Pauli spin matrix for the selected two-level NV transition, while $\hat{a}_{j}^{\dag }, \hat{a}_{j}$ are the creation and annihilation operators for the QMF mode in arm $j$. The nuclear bath couples to the NV-electron spin through the term
\begin{eqnarray}
\hat{H}_{EB,j}=\hat{\sigma}_{z,j}\sum_{k=1}^{N_j} \left [ A(\vec{r}_k)\hat{\tau}_{k,z}+B(\vec{r}_k)\left ( \hat{\tau}_{k,x}{\rm cos}\phi_k+\hat{\tau}_{k,y}{\rm sin}\phi_k \right ) \right ], \nonumber\\
\label{Eq:m4}
\end{eqnarray}
where $\hat{\tau}_{k,x}$ and $\hat{\tau}_{k,y}$ denote the Pauli spin operators for the nuclear spin. The unit vector joining the electron and the {\it k-th} nuclear spin $\vec{r}_k=(r_k,\theta_k,\phi_k)$ is characterized by the polar angle $\theta_k$ and azimuthal angle $\phi_k$ and $N_{j}$ is the number of $^{13}C$ nuclear spin in the $j$-th diamond lattice.

The large difference between electron and nuclear Zeeman energies leads to ignore flip-flop terms involving $\hat{\sigma}_x$ and $\hat{\sigma}_y$ operators.
The coupling strengths in Eq.(\ref{Eq:m4}) are
\begin{eqnarray}
A(\vec{r}_k)=-\frac{\mu_0}{4\pi}\frac{\gamma_{NV}\gamma_C}{r_k^3}\left [ 3{\rm cos}^2(\theta_k)-1 \right ],
\label{Eq:m5}
\end{eqnarray}
and
\begin{eqnarray}
B(\vec{r}_k)=-\frac{\mu_0}{4\pi}\frac{\gamma_{NV}\gamma_C}{r_k^3} 3{\rm cos}(\theta_k){\rm sin}(\theta_k),
\label{Eq:m6}
\end{eqnarray}
where $\gamma_{NV}$ ($\gamma_C$) denotes the gyromagnetic ratio of the NV electron (nuclear) spin and $r_k$ is the distance between the NV and the {\it k-th} nucleus in the diluted spin bath.

The local nuclear spin bath Hamiltonian, $\hat{H}_{B,j}$, is given by:
\begin{equation}
\hat{H}_{B,j}=\hat{H}_{N,j}+\hat{H}_{DD,j},
\end{equation}
where
\begin{eqnarray}
\hat{H}_{N,j}=\sum_{k=1}^{N_j}\frac{\omega_{k}}{2}\hat{\tau}_{k,z},
\label{Eq:m2}
\end{eqnarray}
and
\begin{eqnarray}
\hat{H}_{DD,j}=\sum_{i< k}C_{i,k}\left ( 3\hat{\tau}_{i,z}\hat{\tau}_{k,z}-\hat{\overrightarrow{\tau}}_i\cdot \hat{\overrightarrow{\tau}}_k \right ),
\label{Eq:m3}
\end{eqnarray}
with $\omega_{k}$ the Zeeman energy splitting for the nuclear bath spins and the intrabath secular dipolar coupling strengths are given by
\begin{eqnarray}
C_{i,k}=-\frac{\mu_0}{4\pi}\frac{\gamma_C^2}{r_{i,k}^3}\left [ 3{\rm cos}^2(\theta_{i,k})-1 \right ],
\label{Eq:m7}
\end{eqnarray}
and $r_{i,k}$ denotes the distance between nuclei $i$ and $k$, while $\theta_{i,k}$ is the polar angle formed by the unit vector joining these two bath nuclei and the $z$-direction.

Exact solution of the dynamics for the system described by Eq.(\ref{Eq:m1}) implies a huge number of correlations between the central spin, the QMF and the nuclear spin bath. We propose an alternative solution for the problem: we approximate the spin $^{13}C$ bath with a classical noise field acting on the central
spin ~\cite{Wang2012}. Therefore, let us denote by
\begin{eqnarray}
\hat{H} &=&\sum_{j=A,B}\left[ \frac{\omega _{j}}{2}\hat{\sigma}
_{j,z}+\Omega _{j}\hat{a}_{j}^{\dag }\hat{a}_{j}+G_{j}(t)\left( \hat{a}
_{j}^{\dag }\hat{\sigma}_{j}^{-}+\hat{a}_{j}\hat{\sigma}_{j}^{+}\right)
\right.].\nonumber
\\
\label{Eqq:m8}
\end{eqnarray}
the effective Hamiltonian coming from Eq.(\ref{Eq:m1}). The spin bath terms $\hat{H}_{EB,j}$, $\hat{H}_{N,j}$ and $\hat{H}_{DD,j}$ have been approximated as a time dependent electron-cavity coupling $G_{j}(t)$ implementing a proposal for a classical field in our work ~\cite{Dobrovitski2009}. The long-range character of the dipolar coupling between $%
^{13}C$ nuclear bath spins warrants this
approximation. More concretely, such a noise field is represented by an
Ornstein-Uhlenbeck random process, which is Gaussian and
stationary. In order to guarantee these conditions we consider the following elements:
First, due to the long-range character of the dipolar coupling, the NV
experiences the action of a large number of the bath spins with comparable
strength, therefore this field can be modeled as a Gaussian field with zero
mean. Second, due to the interaction between a single NV and a $^{13}C$ ($A(\vec{r}_{k})\approx B(\vec{r}_{k})=10\ KHz-50\ KHz$) is small in comparison with the action of hundred of spin bath on the NV ($C_{i,k}\approx 2KHz-10KHz$) we can assume a small back action and satisfies the
stationary condition.
Since now, for include the noise environment we consider the case of a stochastic term added to the constant $g_{j}$,
i.e. $G_{j}(t)=g_{0,j}+g_{j}(t)$ ~\cite{Dobrovitski2009}. In the simplest case, when no nuclear bath is affecting the NV-electron dynamics, we retrieve a coupling term constant $G_{j}(t)=g_{0,j}$. The stochastic term $g_{j}(t)$ is described by an Ornstein-Uhlenbeck stochastic process defined by its moments ~\cite{Wang2012}
\begin{eqnarray}
\left\langle g_{j}(t)\right\rangle &=&0,
\label{p7}
\end{eqnarray}%
\begin{eqnarray}
\left\langle g_{j}(t)g_{j^{\prime}}(t^{\prime })\right\rangle &=&b_j^{2}e^{-\frac{%
\left\vert t-t^{\prime }\right\vert }{\tau_j}}\delta_{j,j^{\prime}}.
\label{p8}
\end{eqnarray}%
The dispersion $b_j$ depends on the coupling between the central spin and the spin bath, while the correlation decay rate $\tau_j$ is determined by the intra-coupling among nuclear spins of the $j$-th bath. For a justification of a similar Hamiltonian in a classical context (no microwave photons but classical microwave pulses in a rotating frame) and single NV-bath system see \cite{Slichter,Hanson2006,Dobrovitski2009}.

Now, we proceed to analyze the effect of this noisy environment on the entanglement dynamics for the NV electronic spins.
We consider as initial state two NV electronic spins in their ground states $\left\vert e_{g}\right\rangle_{A}$, $\left\vert e_{g}\right\rangle_{B}$
and the field in a two mode squeezed state $\left\vert r\right\rangle$
\begin{equation}
\left\vert r\right\rangle =\frac{1}{\cosh \left( r\right) }%
\sum_{n=0}^{\infty }\tanh ^{n}(r)\left\vert n,n\right\rangle _{A,B},
\label{eqa:14}
\end{equation}
where the state for the radiation can be understood as the superposition of twin photons
propagating on spatially separated transmission lines ~\cite{Flurin12}. The parameter $r$ in Eq.~\eqref{eqa:14} is the squeezing value for the field and determines the degree of entanglement of the QMF ~\cite{Gauss1}. The photon number in modes $A$ and $B$ are indicated with $n$ in Eq.~\eqref{eqa:14}.
Now we proceed to evaluate the reduced two-spin density operator.
First, we calculate the state of the system at time $t$, $\left\vert \psi(t) \right\rangle$ (see the Appendix \ref{sec:appendixa}), then
we obtain tracing over the photon states $p$ and $q$ (see details in Appendix \ref{sec:appendix1} for a full density matrix expression $\bar{\rho}(t)$)
\begin{equation}
\bar{\rho}_{2e}(t)=\sum_{p=0}^{\infty }\sum_{q=0}^{\infty }\left\langle
p,q\right\vert \bar{\rho}(t)\left\vert p,q\right\rangle,
\end{equation}
yielding to
\begin{equation}
\bar{\rho}_{2e}(t)=\left(
\begin{array}{cccc}
\rho _{1,1}(t) & 0 & 0 & \rho _{1,4}(t) \\
0 & \rho _{2,2}(t) & 0 & 0 \\
0 & 0 & \rho _{3,3}(t) & 0 \\
\rho _{4,1}(t) & 0 & 0 & \rho _{4,4}(t)%
\label{ecrho2}
\end{array}%
\right),
\end{equation}
where the bar in $\bar{\rho_{2e}}(t)$ denotes averages over any stochastic term affecting the spin-cavity coupling term. The non-zero diagonal matrix elements of $\bar{\rho_{2e}}(t)$ are
\begin{widetext}
\begin{eqnarray}
\rho _{1,1}(t)=\frac{1}{r_{c}^{2}}\sum_{n=0}^{\infty }\frac{r_{t}^{2n}}{16}%
\left( 2+\left\langle e^{2i\theta _{A,n}(t)}\right\rangle +\left\langle
e^{-2i\theta _{A,n}(t)}\right\rangle \right) \left( 2+\left\langle
e^{2i\theta _{B,n}(t)}\right\rangle +\left\langle e^{-2i\theta
_{B,n}(t)}\right\rangle \right),\nonumber\\
\rho _{2,2}(t)=\frac{1}{r_{c}^{2}}\sum_{n=0}^{\infty }\frac{r_{t}^{2n}}{16}%
\left( 2+\left\langle e^{2i\theta _{A,n}(t)}\right\rangle +\left\langle
e^{-2i\theta _{A,n}(t)}\right\rangle \right) \left( 2-\left\langle
e^{2i\theta _{B,n}(t)}\right\rangle -\left\langle e^{-2i\theta
_{B,n}(t)}\right\rangle \right),\nonumber\\
\rho _{3,3}(t)=\frac{1}{r_{c}^{2}}\sum_{n=0}^{\infty }\frac{r_{t}^{2n}}{16}%
\left( 2-\left\langle e^{2i\theta _{A,n}(t)}\right\rangle -\left\langle
e^{-2i\theta _{A,n}(t)}\right\rangle \right) \left( 2+\left\langle
e^{2i\theta _{B,n}(t)}\right\rangle +\left\langle e^{-2i\theta
_{B,n}(t)}\right\rangle \right),\nonumber\\
\rho _{4,4}(t)=\frac{1}{r_{c}^{2}}\sum_{n=0}^{\infty }\frac{r_{t}^{2n}}{16}%
\left( 2-\left\langle e^{2i\theta _{A,n}(t)}\right\rangle -\left\langle
e^{-2i\theta _{A,n}(t)}\right\rangle \right) \left( 2-\left\langle
e^{2i\theta _{B,n}(t)}\right\rangle -\left\langle e^{-2i\theta
_{B,n}(t)}\right\rangle \right),
\label{Eqq3}
\end{eqnarray}
\end{widetext}
while the non-diagonal elements are%
\begin{widetext}
\begin{eqnarray}
\rho _{1,4}(t) &=&\rho _{4,1}^{\ast }(t) \nonumber\\
&=&\frac{1}{r_{c}^{2}}\sum_{n=0}^{\infty }\frac{r_{t}^{2n+1}}{16}\left(
\left\langle e^{i\left( \theta _{A,n}(t)-\theta _{A,n+1}(t)\right)
}\right\rangle -\left\langle e^{i\left( \theta _{A,n}(t)+\theta
_{A,n+1}(t)\right) }\right\rangle +\left\langle e^{-i\left( \theta
_{A,n}(t)+\theta _{A,n+1}(t)\right) }\right\rangle -\left\langle e^{-i\left(
\theta _{A,n}(t)-\theta _{A,n+1}(t)\right) }\right\rangle \right)\nonumber\\
&&\times \left( \left\langle e^{i\left( \theta _{B,n}(t)-\theta
_{B,n+1}(t)\right) }\right\rangle -\left\langle e^{i\left( \theta
_{B,n}(t)+\theta _{B,n+1}(t)\right) }\right\rangle +\left\langle e^{-i\left(
\theta _{B,n}(t)+\theta _{B,n+1}(t)\right) }\right\rangle -\left\langle
e^{-i\left( \theta _{B,n}(t)-\theta _{B,n+1}(t)\right) }\right\rangle
\right),
\label{Eqq4}
\end{eqnarray}
\end{widetext}
where angular brackets in terms such as $\left\langle e^{-i\theta
_{j,n}(t)}\right\rangle $ in Eq.~\eqref{Eqq3}-\eqref{Eqq4}
denote averages over the stochastic process simulating the nuclear bath noise.
Expressions such as $\left\langle e^{-i\int_{0}^{t}g_{r}(t)}\right\rangle $ can be evaluated in a closed form with the noise functions given in Eq.~\eqref{p7}-\eqref{p8}
\begin{equation}
\left\langle e^{-i\int_{0}^{t}g_{r}(t)}\right\rangle =e^{-b^{2}\tau \left[
t+\tau \left( e^{-\frac{t}{\tau }}-1\right) \right] },
\label{Eq51}
\end{equation}%
For future use, the noise effects are summarized in the function%
\begin{equation}
R\left( t,p,b,\tau \right) =e^{-p^{2}b^{2}\tau ^{2}\left[ e^{-\left( \frac{t%
}{\tau }\right) }+\frac{t}{\tau }-1\right] }.  \label{Eq53}
\end{equation}%

In order to test the validity of the above expressions we have considered the case where the spin-cavity couplings are constant and identical, i.e. $%
g_{0,A}(t)=g_{0,B}(t)=g$, thus no-stochastic average is required, see Appendix ~\ref{sec:appendixB}.

Expressions given by Eqs.~\eqref{Eqr1}-\eqref{Eqr4} agree perfectly with those reported in ~\cite{Paternostro04}. Additionally, the entanglement transferred from the QMF to the pair of spins is simply obtained as    $\varepsilon
_{NPT}(t)=-2\left[ \rho _{1,4}(t)+\rho _{3,3}(t)\right] $.
Introducing decoherence and dissipation effects and considering each NV spin coupled to its own spin bath with parameters $b_{j}$ and $\tau _{j}$ ($j=A,B$). Besides, identical deterministic spin-cavity constants strengths, $g_{0,A}=g_{0,B}=g$, Eqs.~\eqref{Eqq3}-\eqref{Eqq4} become
\begin{widetext}
\begin{eqnarray}
\rho _{1,1}(t)&=&\frac{1}{r_{c}^{2}}\sum_{n=0}^{\infty }\frac{r_{t}^{2n}}{4}%
\left[ 1+\cos \left( 2\sqrt{n}gt\right) R\left( t,2\sqrt{n},b_{A},\tau
_{A}\right) \right] \left[ 1+\cos \left( 2\sqrt{n}gt\right) R\left( t,2\sqrt{%
n},b_{B},\tau _{B}\right) \right],\nonumber\\
\rho _{2,2}(t)&=&\frac{1}{r_{c}^{2}}\sum_{n=0}^{\infty }\frac{r_{t}^{2n}}{4}%
\left[ 1+\cos \left( 2\sqrt{n}gt\right) R\left( t,2\sqrt{n},b_{A},\tau
_{A}\right) \right] \left[ 1-\cos \left( 2\sqrt{n}gt\right) R\left( t,2\sqrt{%
n},b_{B},\tau _{B}\right) \right],\nonumber\\
\rho _{3,3}(t)&=&\frac{1}{r_{c}^{2}}\sum_{n=0}^{\infty }\frac{r_{t}^{2n}}{4}%
\left[ 1-\cos \left( 2\sqrt{n}gt\right) R\left( t,2\sqrt{n},b_{A},\tau
_{A}\right) \right] \left[ 1+\cos \left( 2\sqrt{n}gt\right) R\left( t,2\sqrt{%
n},b_{B},\tau _{B}\right) \right],\nonumber\\
\rho _{4,4}(t)&=&\frac{1}{r_{c}^{2}}\sum_{n=0}^{\infty }\frac{r_{t}^{2n}}{4}%
\left[ 1-\cos \left( 2\sqrt{n}gt\right) R\left( t,2\sqrt{n},b_{A},\tau
_{A}\right) \right] \left[ 1-\cos \left( 2\sqrt{n}gt\right) R\left( t,2\sqrt{%
n},b_{B},\tau _{B}\right) \right],\nonumber\\
\rho _{1,4}(t) &=&\rho _{4,1}^{\ast } \nonumber\\
&=&-\frac{1}{r_{c}^{2}}\sum_{n=0}^{\infty }\frac{r_{t}^{2n+1}}{4}\times  \nonumber\\
&&\times \left[ \sin \left( \left( \sqrt{n}-\sqrt{n+1}\right) gt\right)
R\left( t,\sqrt{n}-\sqrt{n+1},b_{A},\tau _{A}\right) -\sin \left( \left(
\sqrt{n}+\sqrt{n+1}\right) gt\right) R\left( t,\sqrt{n}+\sqrt{n+1}%
,b_{A},\tau _{A}\right) \right] \times  \nonumber\\
&&\left[ \sin \left( \left( \sqrt{n}-\sqrt{n+1}\right) gt\right) R\left( t,%
\sqrt{n}-\sqrt{n+1},b_{B},\tau _{B}\right) -\sin \left( \left( \sqrt{n}+%
\sqrt{n+1}\right) gt\right) R\left( t,\sqrt{n}+\sqrt{n+1},b_{B},\tau
_{B}\right) \right].
\label{pe1}
\end{eqnarray}
\end{widetext}
From Eqs.~\eqref{pe1} it is straightforward to derive the degree of entanglement between two electronic spins including the environmental dynamics of nuclear spin baths surrounding the two central NV systems. To measure the degree of entanglement contained in the spin quantum state we have used the Wootters concurrence\cite{Wootters98}.

\section{QMF power entangling over two distant electron/nuclear spins in noisy NVs}
\label{sec:B}
In the previous section we discussed a simple situation where only the electronic spin of each NV center have been considered. We are now able to go beyond that simple scenario. More realistically, each NV center is composed of an electronic spin, $e_{j}$, coupled via a hyperfine interaction to a nearest neighbor nuclear spin $\nu_{j}$ ($j=A,B$) which can be that of the substitutional nitrogen atom itself $^{14-15}N$ or a $^{13}C$ atom in the first-shell, see Fig.~\ref{fig:setup}-(b). The Hamiltonian for this system including the nuclear bath within the mean field approximation is
\begin{eqnarray}
\nonumber \hat{H} &=&\hat{H}_A+\hat{H}_B\\
\nonumber &=&\sum_{j=A,B}\left \{ \frac{\omega _{j}}{2}\hat{\sigma}
_{j,z}+\Omega _{j}\hat{a}_{j}^{\dag }\hat{a}_{j}+G_{j}(t)\left( \hat{a}
_{j}^{\dag }\hat{\sigma}_{j}^{-}+\hat{a}_{j}\hat{\sigma}_{j}^{+}\right)
\right.\\
\nonumber &&+\left.
\hat{\sigma}_{z,j}\left [ A(\vec{r}_j)\hat{\tau}_{j,z}+B(\vec{r}_j)\left ( \hat{\tau}_{j,x}{\rm cos}\phi_j+\hat{\tau}_{j,y}{\rm sin}\phi_j \right ) \right ] \right \},\\
\label{Eq:m8}
\end{eqnarray}
where the unit vector joining the electron-nuclear spin pair in the $j$-th NV is given by $\vec{r}_j=(r_j,\theta_j,\phi_j)$ with the polar angle $\theta_j$ and azimuthal angle $\phi_j$, respectively. Expressions for coefficients $A(\vec{r}_j)$ and $B(\vec{r}_j)$ are the same as those quoted in Eqs.~\eqref{Eq:m5}-~\eqref{Eq:m6}. Let us now proceed to analyze the spin pair system's entanglement dynamics. Since subsystems $A$ and $B$ are independent, their respective Hamiltonian
operators commute $\hat{H}=\hat{H}_{A}+\hat{H}_{B}$ with $\left[ \hat{H}_{A},%
\hat{H}_{B}\right] =0$. The Hamiltonian in Eq.(\ref{Eq:m8}) commutes with the total excitation number operator
\begin{equation}
\hat{{\cal N}}=\hat{{\cal N}_{A}}+\hat{{\cal N}_{B}},
\end{equation}
\begin{equation}
\hat{{\cal N}_{j}}=\hat{a}_{j}^{\dag }\hat{a}_{j}+\left( \frac{\hat{\sigma}_{j,z}+1}{2}\right)
+\left( \frac{\hat{\tau}_{j,z}+1}{2}\right); j=A,B.
\label{Eq2}
\end{equation}
Consequently, for each subsystem a sub-space with a well defined number of excitations presents a closed dynamics which proceeds independently from other sub-spaces with different excitation number. Let $\left\vert n,e_{\sigma},\nu _{\tau}\right\rangle_{j} $ denotes a general state for the subsystem $j$ with $n=0,1,2,3,...$ photons, the electron in one of the states $\left\vert e_{\sigma }\right\rangle_{j}
=\left\vert e_{g}\right\rangle_{j} ,\left\vert e_{e}\right\rangle_{j} $ with $\hat{%
\sigma}_{j,z}\left\vert e_{g}\right\rangle_{j} =-\left\vert e_{g}\right\rangle_{j} ,\
\hat{\sigma}_{j,z}\left\vert e_{e}\right\rangle_{j} =\left\vert e_{e}\right\rangle_{j}
$ and the nucleus in state $\left\vert \nu _{\tau }\right\rangle_{j} =\left\vert
\nu _{g}\right\rangle_{j} ,\left\vert \nu _{e}\right\rangle_{j} $ with $\hat{\tau}%
_{j,z}\left\vert \nu _{g}\right\rangle_{j} =-\left\vert \nu _{g}\right\rangle_{j} ,\
\hat{\tau}_{j,z}\left\vert \nu _{e}\right\rangle_{j} =\left\vert \nu
_{e}\right\rangle_{j} $. Thus, the full Hilbert space for each subsystem can be partitioned into independent sub-spaces in the following way: a one-dimensional subspace corresponding to the state $\left\vert 0,e_{g},\nu _{g}\right\rangle_{j} $ with $%
{\cal N}_{j}=0$ excitations; a single three dimensional sub-space, with $N_{j}=1,$ spanned by the vectors
\begin{equation}
\left\vert 1,1\right\rangle_{j} =\left\vert 0,e_{e},\nu _{g}\right\rangle_{j},
\label{Eq3}
\end{equation}%
\begin{equation}
\left\vert 1,2\right\rangle_{j} =\left\vert 1,e_{g},\nu _{g}\right\rangle_{j},
\label{Eq4}
\end{equation}%
\begin{equation}
\left\vert 1,3\right\rangle_{j} =\left\vert 0,e_{g},\nu _{e}\right\rangle_{j},
\label{Eq5}
\end{equation}
and finally an infinite number of four dimensional subspaces with ${\cal N}_{j}\geq 2$ (or equivalently $n\geq 1$ given the fact that ${\cal N}_{j}=n_{j}+1$) spanned by vectors
\begin{equation}
\left\vert N,1\right\rangle_{j} =\left\vert n,e_{e},\nu _{g}\right\rangle_{j},
\label{Eq6}
\end{equation}%
\begin{equation}
\left\vert N,2\right\rangle_{j} =\left\vert n+1,e_{g},\nu _{g}\right\rangle_{j},
\label{Eq7}
\end{equation}%
\begin{equation}
\left\vert N,3\right\rangle_{j} =\left\vert n-1,e_{e},\nu _{e}\right\rangle_{j},
\label{Eq8}
\end{equation}%
\begin{equation}
\left\vert N,4\right\rangle_{j} =\left\vert n,e_{g},\nu _{e}\right\rangle_{j}.
\label{Eq9}
\end{equation}
We assume an unentangled initial state of the form $\left\vert \psi (0)\right\rangle =\left\vert
r\right\rangle \otimes \left\vert e_{g},\nu _{g}\right\rangle _{A}\otimes
\left\vert e_{g},\nu _{g}\right\rangle _{B}$, where the initial state for the microwave radiation has the same form as in Eq.(\ref{eqa:14}). At later times the system's state becomes
\begin{eqnarray}
\left\vert \psi (t)\right\rangle &=&\hat{U}_{A,B}(t)r_{c}\sum_{n=0}^{\infty
}r_{t}^{n}\left\vert n,e_{g},\nu _{g}\right\rangle _{A}\otimes \left\vert
n,e_{g},\nu _{g}\right\rangle _{B}  \nonumber \\
&=&r_{c}\sum_{n=0}^{\infty }r_{t}^{n}\left[ \hat{U}_{A}(t)\left\vert
n,e_{g},\nu _{g}\right\rangle _{A}\right] \otimes \left[ \hat{U}%
_{B}(t)\left\vert n,e_{g},\nu _{g}\right\rangle _{B}\right],\nonumber \\
\label{Eqq58}
\end{eqnarray}
The evolution operator is $\hat{U}_{A,B}(t)=\hat{U}_{A}(t)\otimes \hat{U}%
_{B}(t)$ because we consider independent subsystems. The total evolution operator $\hat{U}_{A,B}$ is determined by the system's
Hamiltonian given by Eq.~\eqref{Eq:m8}. The state in Eq.~\eqref{Eqq58}, can be expandend in terms of a set of time dependent coefficients and the base states Eqs.~\eqref{Eq3}-\eqref{Eq9}
\begin{widetext}
\begin{eqnarray}
\left\vert \psi (t)\right\rangle =\sum_{j=A,B}\left[ C_{0,j}\left\vert
0,e_{g},\nu _{g}\right\rangle _{j}+\sum_{k=1}^{3}C_{1,k}(t)\left\vert
1,k\right\rangle _{j}+\sum_{N=2}^{\infty }\sum_{k=1}^{4}C_{N,k}(t)\left\vert
N,k\right\rangle _{j}\right].
\label{pt11}
\end{eqnarray}
\end{widetext}
Due to the inclusion of the hyperfine interaction between the $e_{j}-\nu_{j}$ spins we can not obtain analytical expressions for the density matrix that characterize the dynamical evolution of the spin system $\rho(t)=\left\vert \psi (t)\right\rangle \left\langle \psi (t)\right\vert$, therefore we have calculated numerically the density matrix
for the system $\bar{\rho}(t)=\left\vert \psi (t)\right\rangle \left\langle \psi (t)\right\vert$ and then the reduced density operator $\bar{\rho}_{2Q}(t)$ ($16 \times 16$ matrix) tracing over the states of the field
\begin{eqnarray}
\bar{\rho}_{2Q}(t)=\sum_{p=0}^{\infty }\sum_{q=0}^{\infty }\left\langle
p,q\right\vert \bar{\rho}(t)\left\vert p,q\right\rangle,
\label{p11}
\end{eqnarray}
where $p$ and $q$ represent the photon states number in the two branches and the bar in $\bar{\rho}_{2Q}(t)$ and $\bar{\rho}(t)$ denotes averages over the stochastic term affecting the spin-cavity coupling term.

Before starting to use this formalism, we have compared the numerical results in the case where the hyperfine interaction between the $e_{j}-\nu_{j}$ spins is zero, with the analytical expressions obtained in sec.~\ref{sec:A}.
First, we found numerically the term $\left\langle \left\vert C_{1,2}(t)\right\vert ^{2}\right\rangle$, where $C_{1,2}(t)$ is one of the time dependent coefficient in Eq.~\eqref{pt11}, then we compare this solution with the analytical expression
\begin{equation}
\left\langle \left\vert C_{1,2}(t)\right\vert ^{2}\right\rangle =\frac{1}{2}%
\left( 1-\cos \left( 2gt\right) R\left( t,2,b_{A},\tau _{A}\right) \right).
\end{equation}
The above expression was obtained using the analytical result for the density matrix presented in appendix A. (Eq. A1). In Fig.~\ref{figr1}(a) we present the obtained results. The next step, was compare the analytical and numerical density matrix elements, in Fig.~\ref{figr1}(b) we show the result for $\rho_{11}(t)$, and similar results were obtained for the other density matrix entries. Finally, we evaluate the concurrence between two electronic spins $e_{A}-e_{B}$
with the analytical and numerical techniques, the results are shown in Fig.~\ref{figr1}(c). The Fig.~\ref{figr1}(a)-(b)-(c), were realized with noise conditions $b_{A}=b_{B}=0.5g$, $g\tau_{A}=g\tau_{B}=0.5$
and $n=1000$ numerical realizations. This results allow determine the number of realizations where the numerical results converge with the analytical solutions.
Now we are ready to use this formalism, following the procedure described before, to evaluate the photon induced spin quantum correlations. In the stochastic simulation we have considered $10^4$ realizations for assuring numerical convergence in the calculation of these averages.
\begin{figure}[t]
\begin{center}
\includegraphics[width=0.4\textwidth]{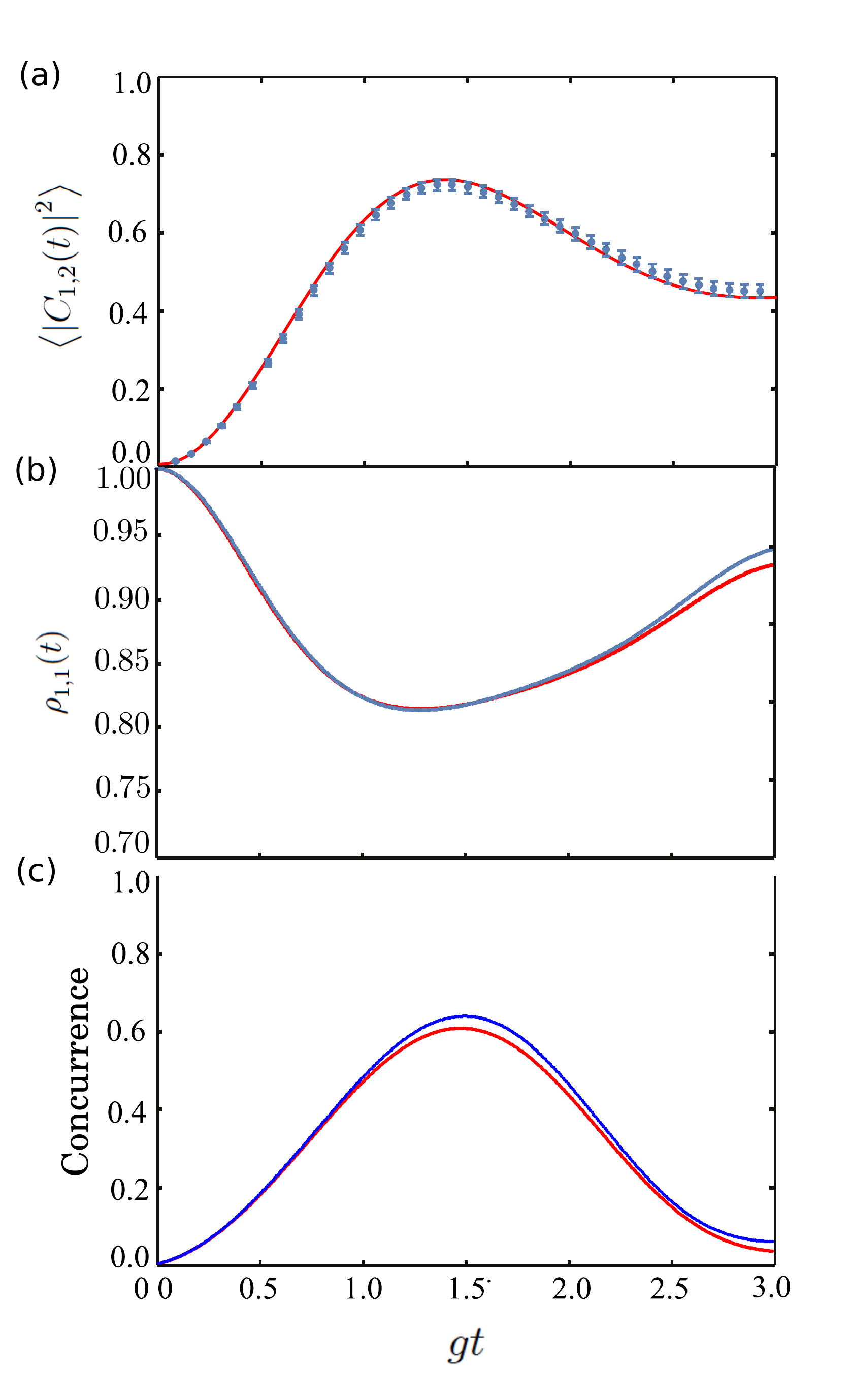}
\end{center}
\caption{Comparison between the analytical and numerical solution in the case where the hyperfine interaction between the $e_{j}-\nu_{j}$ ($j=A,B$) spins is zero. The red line corresponds to the analytical solution, the blue line (points) the numerical results. Noise conditions were included with $b_{A}=b_{B}=0.5g$, $g\tau_{A}=g\tau_{B}=0.5$ and $n=10^3$ numerical realizations.(a) $\left\langle \left\vert C_{1,2}(t)\right\vert ^{2}\right\rangle$ coefficient associated to two electronic spins as a function of time (in g units) with the corresponding error bars.
(b) Density matrix entrie $\rho_{11}(t)$ of two electronic spins as a function of time (in g units), with $r=0.5$. (c) Concurrence between two electronic spins as a function of time (in g units), for $r=0.5$.}
\label{figr1}
\end{figure}
\subsection{Non-local electron-electron ($e_{A}-e_{B}$) entanglement}
\label{subsec:theory2}
In this section we calculate the entanglement between $e_{A}-e_{B}$ spins under noise conditions, including the hyperfine interaction between the nuclear spins associated to each NV center ($\nu_{A}$ and $\nu_{B}$). In this situation Eqs.~\eqref{pe1} are not valid, therefore we require to return to Eqs.~\eqref{p11} and calculate the reduced density matrix. Due to the inclusion of the nuclear spin interaction we cannot anymore evaluate analytically the density matrix entries, therefore we require to evaluate them numerically with an appropriate average over many realizations of the noise effects. In a four-spin base, ordered as $\left\{ \left\vert e_{g},\nu _{g}\right\rangle
,\left\vert e_{g},\nu _{e}\right\rangle ,\left\vert e_{e},\nu
_{g}\right\rangle ,\left\vert e_{e},\nu _{e}\right\rangle \right\}
_{A}\otimes \left\{ \left\vert e_{g},\nu _{g}\right\rangle ,\left\vert
e_{g},\nu _{e}\right\rangle ,\left\vert e_{e},\nu _{g}\right\rangle
,\left\vert e_{e},\nu _{e}\right\rangle \right\} _{B}$, we obtain a $%
16\times 16$ density matrix. Now it is possible to obtain the two-electron spin reduced density matrix. In the base ordered as $\left\{ \left\vert e_{A,g},
e_{B,g}\right\rangle ,\left\vert e_{A,g},e_{B,e}\right\rangle ,\left\vert
e_{A,e},e_{B,g}\right\rangle ,\left\vert e_{A,e},e_{B,e}\right\rangle
\right\} $, reads as%
\begin{eqnarray}
e(t)=\left(
\begin{array}{cccc}
e_{1,1}(t) & 0 & 0 & e_{1,4}(t) \\
0 & e_{2,2}(t) & 0 & 0 \\
0 & 0 & e_{3,3}(t) & 0 \\
e_{4,1}(t) & 0 & 0 & e_{4,4}(t)%
\end{array}%
\right),
\label{p1}
\end{eqnarray}
with
\begin{eqnarray}
e_{1,1}(t)&=&\rho _{1,1}(t)+\rho _{2,2}(t)+\rho _{5,5}(t)+\rho _{6,6}(t),\nonumber\\
e_{2,2}(t)&=&\rho _{3,3}(t)+\rho _{4,4}(t)+\rho _{7,7}(t)+\rho _{8,8}(t),\nonumber\\
e_{3,3}(t)&=&\rho _{9,9}(t)+\rho _{10,10}(t)+\rho _{13,13}(t)+\rho _{14,14}(t),\nonumber\\
e_{4,4}(t)&=&\rho _{11,11}(t)+\rho _{12,12}(t)+\rho _{15,15}(t)+\rho_{16,16}(t),\nonumber\\
\label{p5}
\end{eqnarray}
\begin{eqnarray}
e_{1,4}(t) &=&e_{4,1}^{\ast }(t) \nonumber\\
&=&\rho _{1,11}(t)+\rho _{2,12}(t)+\rho _{5,15}(t)+\rho _{6,16}(t).
\label{p6}
\end{eqnarray}
Numerical results for different QMF and noise parameters will be discussed in Sect.\ \ref{sec:C}.

\subsection{Non-local electron-nuclear ($e_{A}-\nu_{B}$) entanglement}
\label{subsec:theory4}
Let us now consider the QMF entangling power over an electron-nuclear spin pair in distant NVs under the effects of separate $^{13}C$ spin baths. In a base ordered as $\left\{ \left\vert e_{A,g},\nu
_{B,g}\right\rangle ,\left\vert e_{A,g},\nu _{B,e}\right\rangle ,\left\vert
e_{A,e},\nu _{B,g}\right\rangle ,\left\vert e_{A,e},\nu _{B,e}\right\rangle
\right\} $ the $A$ electron $B\ $nucleus reduced density matrix reads as%
\begin{eqnarray}
q (t)=\left(
\begin{array}{cccc}
q _{1,1}(t) & 0 & 0 & q _{1,4}(t) \\
0 & q _{2,2}(t) & 0 & 0 \\
0 & 0 & q _{3,3}(t) & 0 \\
q _{4,1}(t) & 0 & 0 & q _{4,4}(t)%
\end{array}%
\right),
\label{eqq1}
\end{eqnarray}
with
\begin{eqnarray}
q _{1,1}(t)&=&\rho _{1,1}(t)+\rho _{3,3}(t)+\rho _{5,5}(t)+\rho _{7,7}(t),\nonumber\\
q _{2,2}(t)&=&\rho _{2,2}(t)+\rho _{4,4}(t)+\rho _{6,6}(t)+\rho _{8,8}(t),\nonumber\\
q _{3,3}(t)&=&\rho _{9,9}(t)+\rho _{11,11}(t)+\rho _{13,13}(t)+\rho_{15,15}(t),\nonumber\\
q _{4,4}(t)&=&\rho _{10,10}(t)+\rho _{12,12}(t)+\rho _{14,14}(t)+\rho_{16,16}(t),\nonumber\\
\label{eqq2}
\end{eqnarray}

\begin{eqnarray}
q _{1,4}(t) &=&q _{4,1}^{\ast }(t) \nonumber\\
&=&\rho _{1,10}(t)+\rho _{3,12}(t)+\rho _{5,14}(t)+\rho _{7,16}(t).
\label{eqq3}
\end{eqnarray}

Specific form for the density matrix elements are presented in the appendix B. In order to analyze the entanglement transfer from the QMF to the $e_{A}-\nu_{B}$ system and in particular investigate in detail its dependence on the noise sources we have evaluated numerically Eqs.~\eqref{eqq2}-\eqref{eqq3} with averages over the noise realizations.

\subsection{Non-local nuclear-nuclear ($\nu_{A}-\nu_{B}$) entanglement}
\label{subsec:theory3}
In the previous section we show the mechanism to generate entangled states between electronic spins with a correlated field. Now we investigate the most intriguing possibility of a controlled entanglement generation in a nuclear spin pair in separate NV centers in the diamond lattice. Due to the weak coupling between the correlated field and the nuclear spins, we will use the hyperfine interaction between the electronic and nuclear spins as a mediator of the correlation or quantum bus, this kind of mechanism has been proposed in past to connect a finite number of nuclear spins $I=1/2$ ~\cite{Mehring2006}, and nuclear qubits in NV centers have been coupled employing the magnetic dipole-dipole interaction with electron spins ~\cite{Bermudez2011} . First we obtained the nucleus-nucleus density matrix in a base ordered as $\left\{ \left\vert \nu _{A,g},\nu
_{B,g}\right\rangle ,\left\vert \nu _{A,g},\nu _{B,e}\right\rangle
,\left\vert \nu _{A,e},\nu _{B,g}\right\rangle ,\left\vert \nu _{A,e},\nu
_{B,e}\right\rangle \right\} $ the nucleus-nucleus density matrix reads as
\begin{eqnarray}
\nu(t)=\left(
\begin{array}{cccc}
\nu _{1,1}(t) & 0 & 0 & \nu _{1,4}(t) \\
0 & \nu _{2,2}(t) & 0 & 0 \\
0 & 0 & \nu _{3,3}(t) & 0 \\
\nu _{4,1}(t) & 0 & 0 &  \nu _{4,4}(t)%
\end{array}%
\right),
\label{p20}
\end{eqnarray}
with%
\begin{eqnarray}
\nu _{1,1}(t)&=&\rho _{1,1}(t)+\rho _{3,3}(t)+\rho _{9,9}(t)+\rho
_{11,11}(t),\nonumber\\
\nu _{2,2}(t)&=&\rho _{2,2}(t)+\rho _{4,4}(t)+\rho _{10,10}(t)+\rho
_{12,12}(t),\nonumber\\
\nu _{3,3}(t)&=&\rho _{5,5}(t)+\rho _{7,7}(t)+\rho _{13,13}(t)+\rho
_{15,15}(t),\nonumber\\
\nu _{4,4}(t)&=&\rho _{6,6}(t)+\rho _{8,8}(t)+\rho _{14,14}(t)+\rho
_{16,16}(t),\nonumber\\
\label{p21}
\end{eqnarray}

\begin{eqnarray}
\nu _{1,4}(t) &=& \nu _{4,1}^{\ast }(t)\nonumber \\
&=&\rho _{1,6}(t)+\rho _{3,8}(t)+\rho _{9,14}(t)+\rho _{11,16}(t),
\label{p22}
\end{eqnarray}
in the Appendix B we show the expressions for  Eqs.~\eqref{p21}-\eqref{p22}. We have evaluated numerically the expressions Eqs.~\eqref{p21}-\eqref{p22} for determining each of the density matrix entries.

\section{Results and discussion}
\label{sec:C}

Up to now, we have described the general theoretical formalism necessary for addressing the entanglement transfer from two-mode microwave squeezed radiation to a bipartite system
composed of electronic and/or nuclear spins of spatially separated NV centers. Before going to the discussion of our results,
it is important to assess the point concerning realistic numbers for the NV-microwave coupling strength
to which we turn now our attention by briefly reviewing different proposed setups.
Direct magnetic coupling between an ensemble of NVs and transmission line resonators (TLR)
has been experimentally achieved in the linear or Gaussian regime ~\cite{Amsuss,Wach}, confirming additionally the scaling of the collective coupling strength with the square root of the number of emitters. The reported value for the collective coupling constant between an ensemble of $~10^{12}$ NV centers and the
TLR can attain values up to $g_{col}/2\pi\approx 10 MHz$. Furthermore, the possibility of reaching
strong coupling between individual NV electronic spins and TLR, $g/2\pi\approx 0.1 MHz$, has been analyzed for the case of an interconnecting quantum system such as a nanomechanical resonator ~\cite{Cheng}. Moreover, a closely related method extended those possibilities for reaching strong coupling between a single
NV electronic spin and a TLR ~\cite{Twamley}. In addition, related works have proposed a direct coupling between
NVs and superconducting flux qubits with a coupling of $g/2\pi\approx 12MHz$ for a
NV diamond located at the center of the superconducting small loop ~\cite{Marcos}, and the transfer of single excitations
between the NV ensemble with a flux qubit has also been presented in ~\cite{Munro}. Finally, the strong coupling
between NV qubits and superconducting resonators
has made possible the transfer of quantum states between them, under conditions of a coupling strength on the order of $g/2\pi\approx 10 MHz$ as discussed in ~\cite{Stevepra}.
We stress that the plots we describe below are given in terms of dimensionless quantities
(for instance $gt$ for dimensionless time, among others). So that a feature in the entanglement evolution seen at dimensionless $gt=1$ means approximately occurring at a time $t\sim 10^{-1}-1$ $\mu$s, well within the experimental reach of most of the previously quoted works. Thus,
our general results may be testable under realistic experimental conditions.

In this section we provide additional analysis of the entanglement transfer in the three bipartite systems presented before: electron-electron, electron-nucleus and nucleus-nucleus. In particular we investigate in detail its dependence on the noise sources. The experimental values considered in our calculations are: a driven microwave frequency  $\Omega_{1}=\Omega_{2}$ resonant with the electron spin frequencies $\omega_{1}=\omega_{2}=(3/5)\times 10^{4} g$.

\begin{figure}[t]
\begin{center}
\includegraphics[width=0.5\textwidth]{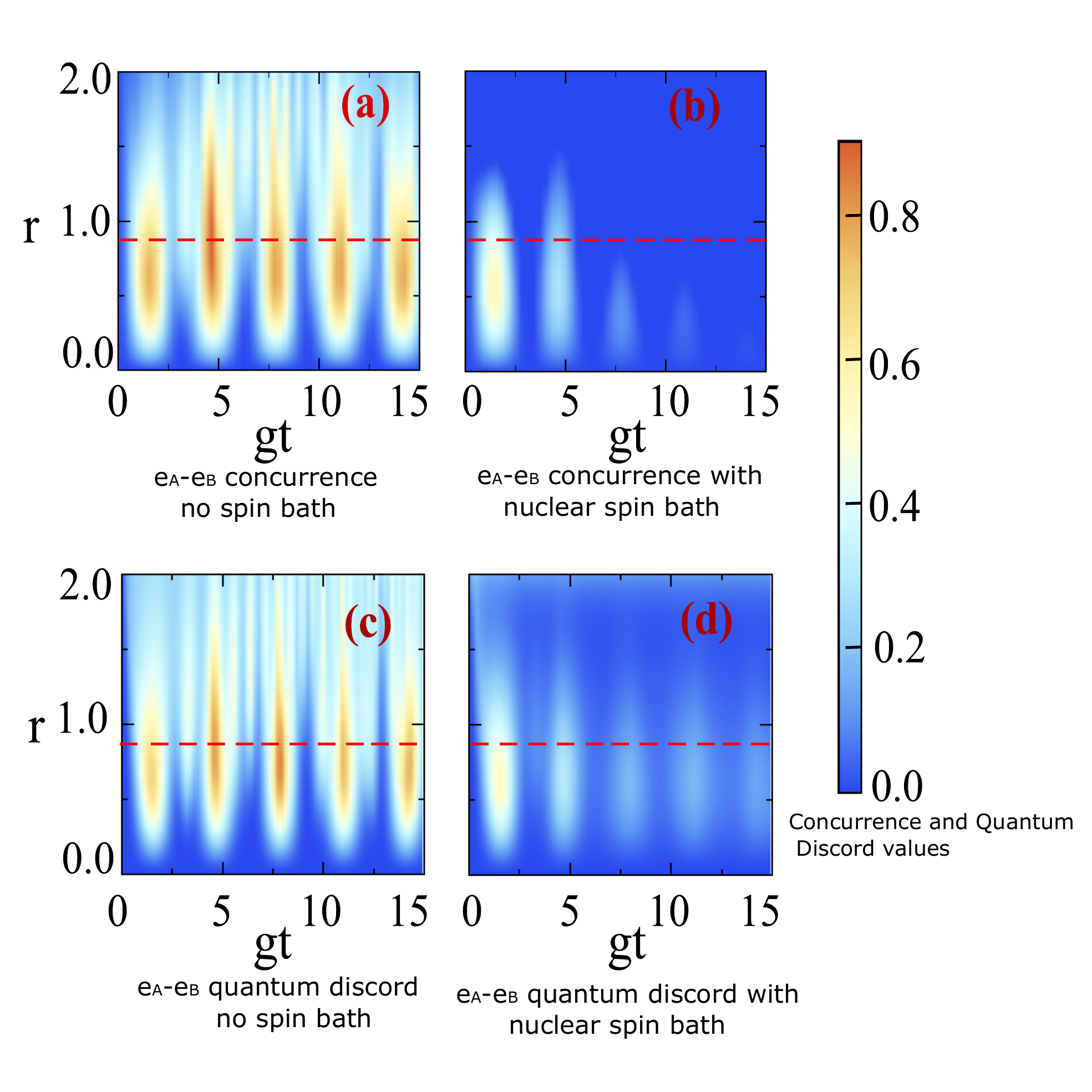}
\end{center}
\caption{$e_{A}-e_{B}$ concurrence (panels (a) and (b)) and quantum discord (panels (c) and (d)) as a function of the QMF squeezing parameter $r$ and dimensionless time $gt$. No nuclear spin bath effects in panels (a) and (c) while nuclear spin effects are displayed in panels (b) and (d) with $g\tau_{A}=g\tau_{B}=0.5$. In all plots the static QMF-spin coupling strength is $g_{0,A}=g_{0,B}=g$.}
\label{fig2}
\end{figure}

\begin{figure}[t]
\begin{center}
\includegraphics[width=0.48\textwidth]{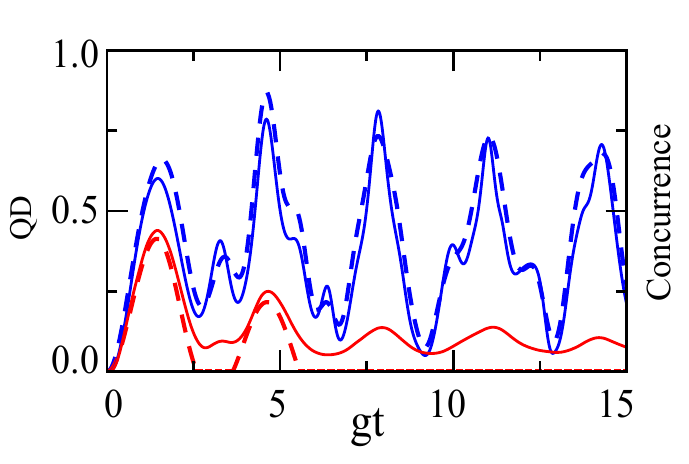}
\end{center}
\caption{$e_{A}-e_{B}$ concurrence (dashed lines) and quantum discord (continuous lines) with symmetric conditions as a function of $gt$ for a selected QMF squeezing parameter $r=0.87$, marked with red dashed lines in Fig.~\ref{fig2}. Blue lines correspond to no nuclear spin bath effects (Fig.~\ref{fig2}-(a,c)) while red lines represent results with nuclear spin bath effects (Fig.~\ref{fig2}-(b,d)) as characterized by $b=0.5g$ and $g\tau=0.5$.}
\label{fig3}
\end{figure}

\begin{figure}[t]
\begin{center}
\includegraphics[width=0.55\textwidth]{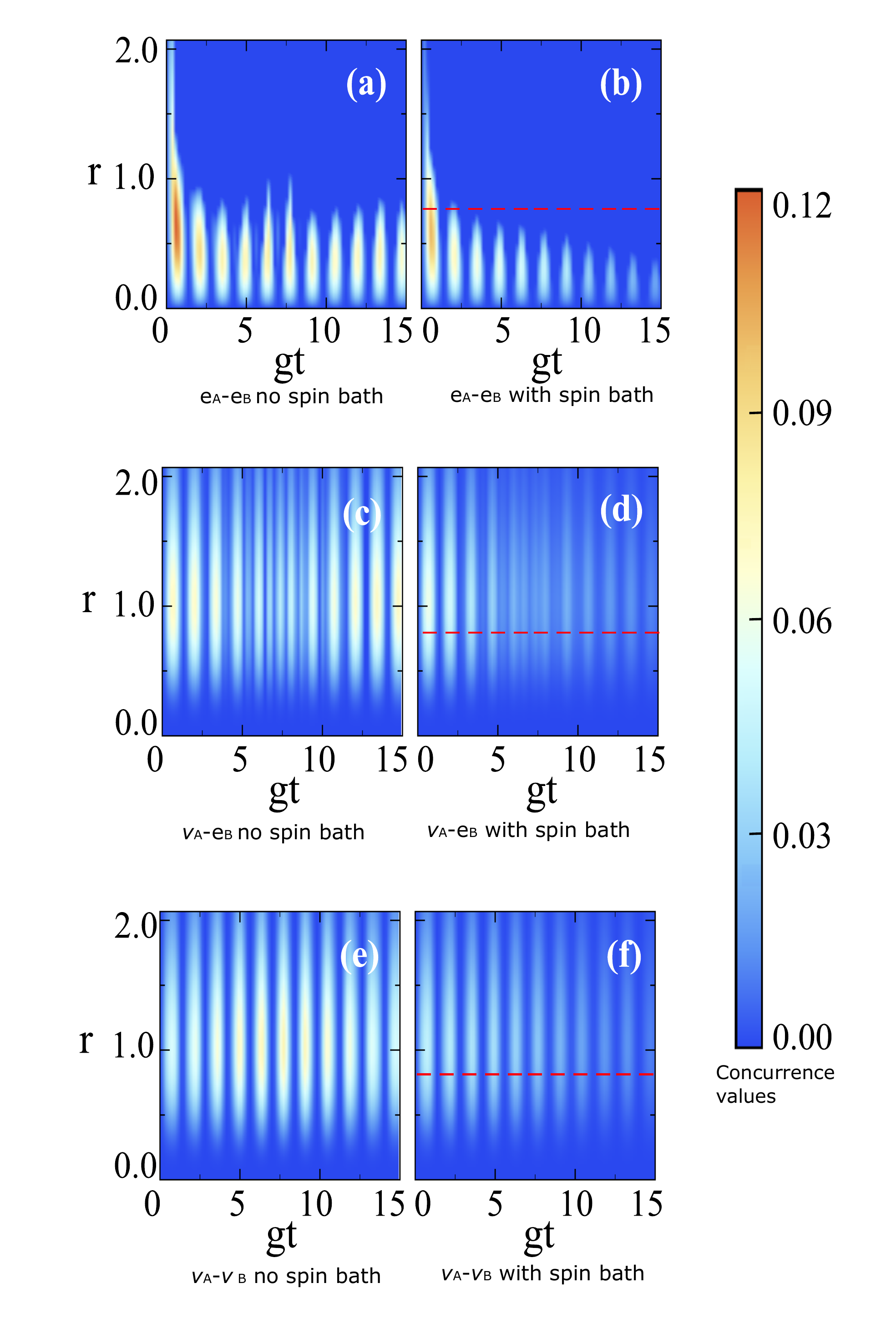}
\end{center}
\caption{Concurrence of different spin pairs in separate NVs as a function of the QMF squeezing parameter $r$ and dimensionless time $gt$. Panels (a) and (b) denote $e_{A}-e_{B}$, panels (c) and (d) represent $e_{B}-\nu_{A}$ (or equivalently $e_{A}-\nu_{B}$), panels (e) and (f) correspond to $\nu_{A}-\nu_{B}$. In all plots the static QMF-spin coupling strength is fixed to $g_{0,A}=g_{0,B}=g$. No nuclear bath effects yield to results in (a), (c) and (e). Nuclear bath effects with $b_{A}=b_{B}=0.5 g$ and $g\tau_{A}=g\tau_{B}=0.5$ in plots (b), (d) and (f).}
\label{fig5}
\end{figure}

\begin{figure}[t]
\begin{center}
\includegraphics[width=0.45\textwidth]{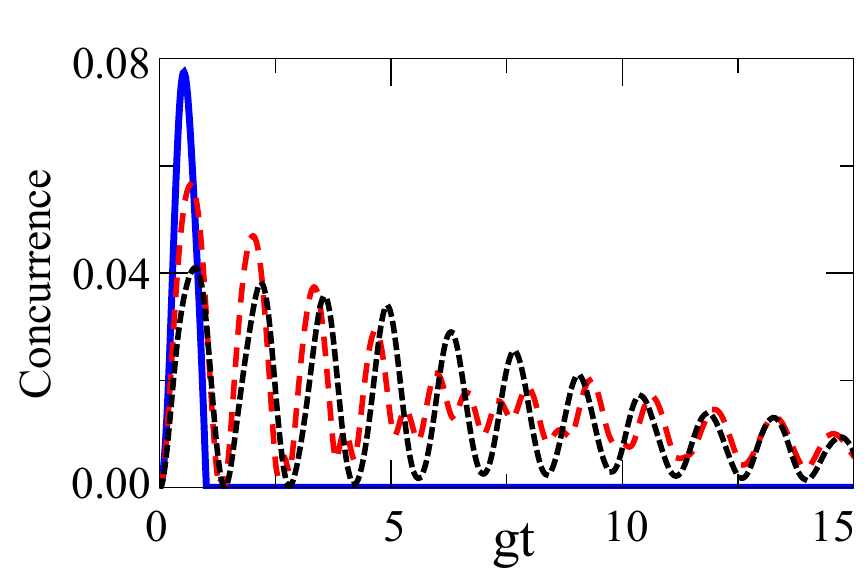}
\end{center}
\caption{Concurrence for distant NV spins as a function of dimensionless time $gt$, symmetric case $g_{0,A}=g_{0,B}=g$, for a selected QMF squeezing parameter $r=0.87$ (marked by dashed red lines in Fig.~\ref{fig5}-(b),(d),(f)). The continuous blue line represents the $e_{A}-e_{B}$ spin pair, the dashed red line $e_{A}-\nu_{B}$ (or ($e_{B}-\nu_{A}$)) and the black dashed line $\nu_{A}-\nu_{B}$.}
\label{fig6}
\end{figure}

\begin{figure}[t]
\begin{center}
\includegraphics[width=0.48\textwidth]{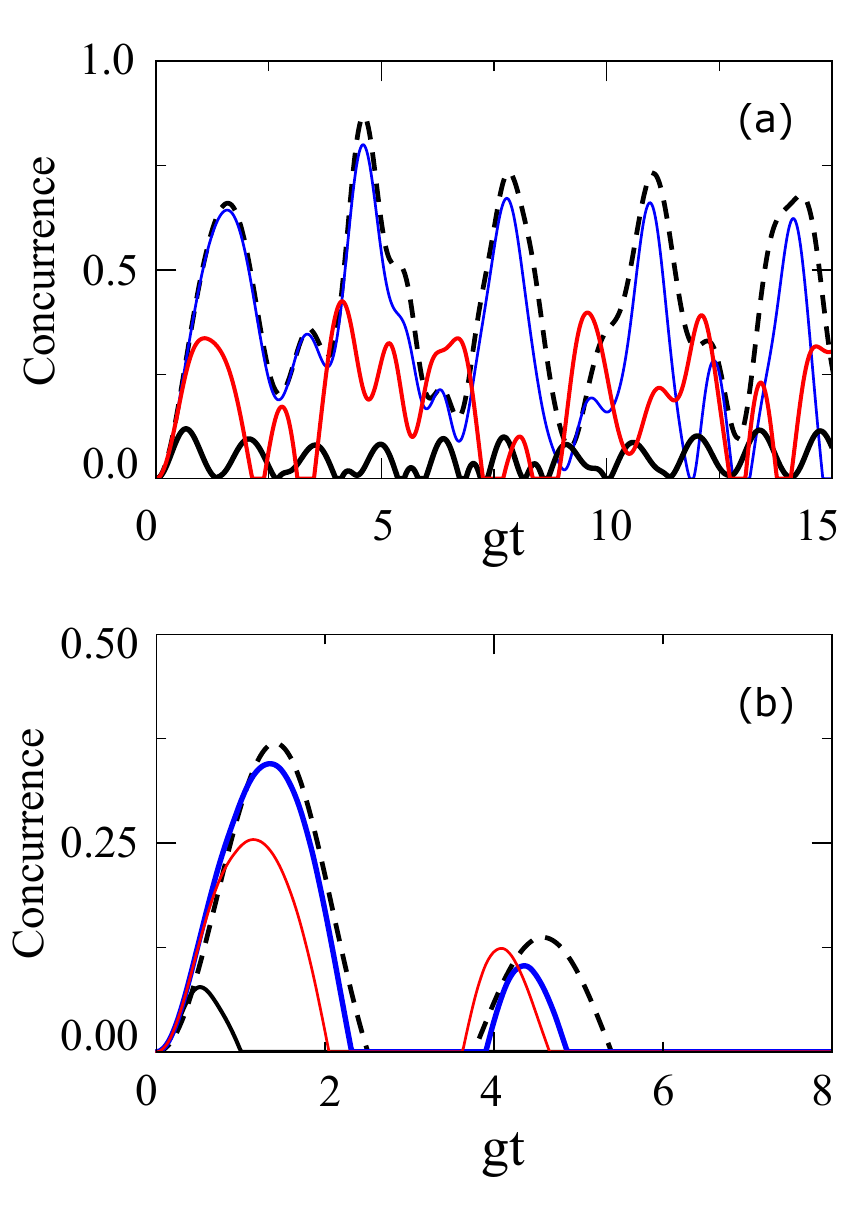}
\end{center}
\caption{$e_{A}-e_{B}$ concurrence as a function of dimensionless time $gt$, symmetric case $g_{0,A}=g_{0,B}=g$ and QMF squeezing parameter $r=0.87$, for selected values of the hyperfine interaction between $e_{j}-\nu_{j}$ in the local $j$-th NV.  Solid lines represent the numerical solution: the small blue line corresponds to $I=0.1 g$ , the red medium line is for $I=1 g$ while the black large line for $I=2 g$. The dashed line represents the exact analytical solution for the case where no hyperfine interaction. (a) No nuclear spin baths. (b) Nuclear spin baths with symmetric noise parameters $b=0.5 g$ and $g\tau=0.5$.
}
\label{fig7}
\end{figure}

As shown in Fig.~\ref{fig2}, we start using the formalism presented in Sect.\ \ref{sec:A} where the hyperfine coupling between the $e_{i}-\nu_{j}$ ($i,j=A,B$) spins $A(\vec{r})=B(\vec{r})= 0$ and we calculate analytically the concurrence and quantum discord between the $e_{A}-e_{B}$ spins. We have plotted two cases: in Fig.~\ref{fig2}(a) we evidence the effective entanglement for the $e_{A}-e_{B}$ spins as function of the squeezing parameter $r$ and time with symmetric conditions for the two branches, $g_{0,A}=g_{0,B}=g$. Furthermore, as the coherent dynamics of the NV centers is strongly influenced by the coupling with neighboring spins ($^{13}C$ spin bath) the noise effect in entanglement transfer simulated with the parameters $b$ and $\tau$ is shown in Fig.~\ref{fig2}(b). Comparing the results between isolated spins Fig.~\ref{fig2}(a) and the realistic situation of the spin bath Fig.~\ref{fig2}(b) we observe a wide region of strong entanglement even with the noisy conditions. As a consequence of the spins bath we note a decrease in the concurrence but principally for large $r$ values. An appreciable entanglement is obtained for $r\leq 1.0$ in both situations: isolated spins and with a spin bath, this value corresponds to a gain $G_{E}=$cosh$^{2}[r] = 2.38$ dB, therefore the required squeezing for the microwaves to obtain
maximum entangled values is in the range of the reported experimental values ~\cite{CastellanosBeltran08}. The results reported allow to determine the optimal region for achieve entanglement in presence of a spin bath. In order to gain insight in the quantum correlations beyond entanglement, we have calculated the quantum discord ~\cite{Ollivier2001} as a function of $r$ and time $gt$ between $e_{A}-e_{B}$ spins  Fig.~\ref{fig2}(c) without spin bath and in a noise environment Fig.~\ref{fig2}(d). Comparing the results between concurrence and quantum discord we can evidence similar behaviors, however the quantum discord persist a longer times while concurrence fall to zero and vanish in the same period of time. A more detailed comparison between concurrence and quantum discord is presented in Fig.~\ref{fig3} where we have selected $r=0.87$ values from Fig.~\ref{fig2} (dashed lines).

Next, we have included the effect of the hyperfine coupling and considered $A(\vec{r})=B(\vec{r})\approx 2 g$. We have evaluated numerically the expressions Eqs.~\eqref{p5}-\eqref{p6}, Eqs.~\eqref{eqq2}-\eqref{eqq3} and Eqs.~\eqref{p21}-\eqref{p22} which include the averages over the coefficients that determine the density matrix entries. In the simulation we have considered $10^4$ stochastic realizations for determining the averages over nuclear noises.

For isolated spin systems the time dependent concurrence is presented in Fig.~\ref{fig5}(a), Fig.~\ref{fig5}(c) and Fig.~\ref{fig5}(e): for $e_{A}-e_{B}$, $e_{A}-\nu_{B}$ and $\nu_{A}-\nu_{B}$, respectively. The bath effect in the entanglement transfer is illustrated in Fig.~\ref{fig5}(b), Fig.~\ref{fig5}(d) and Fig.~\ref{fig5}(f) where the noise parameters are $b=0.5 g$ and $g\tau=0.5$. For the electronic spins the maximum entanglement is achieved for small squeezing value $r$ even including the hyperfine interaction with the proximal nuclear spin. The bath inclusion changes slightly the optimal region to obtain entanglement but small $r$ values are again needed. The dynamics for nuclear spins or the combination of electronic and nuclear spins allows to characterize the strength of the entanglement in terms of the squeezing microwave parameter. The results obtained show that for these systems the amount of squeezing in the microwaves required to produce entanglement is greater compared with the electron pair situation. Beside we can observe that the bath effect is greater in the entanglement between electronic spins, this effect is evidenced more clearly in Fig.~\ref{fig6} where we have selected $r=0.87$ of Fig.~\ref{fig5}(dashed red lines) and evaluate the concurrence as a function of time. The blue continuous line represents the $e_{A}-e_{B}$ entanglement while the medium dashed red line the $e_{A}-\nu_{B}$ and the small dashed black line $\nu_{A}-\nu_{B}$.

It is worth noting that a longtime interest has existed for reaching cross entanglement between different spin species, in special electron-nucleus entanglement, due to the fact of its non-trivial consequences for quantum computing devices. In the field of NMR based quantum information processing, malonic acid molecular single crystals were used to demonstrate that the entanglement between disparate spins (electronic spin resonance in GHz while the nuclear spin resonance is in the frequency domain of MHz) is not only achievable but detectable ~\cite{Mehring03}. On the other hand, magic number transitions in few electron quantum dots have been proposed for affecting and detecting the entanglement between the electron spins and a single nuclear spin, providing reliable quantum gate operations ~\cite{Reina2000}. We stress that results discussed in this section bring an alternative path for reaching such cross entangling, with the added possibility of affecting spatially separated different spin species. 

Finally, the effect of the hyperfine coupling between the $e_{i}-\nu_{j}$ spin is illustrated in Fig.~\ref{fig7} where we compare the exact analytical solution for the concurrence between two electronic spins $e_{A}-e_{B}$ (dashed line) without hyperfine interaction with the numerical solution for $I=0.1 g$, $I=g$ and $I=2 g$, where we have considered $A(\vec{r})=B(\vec{r})=I$. In Fig.~\ref{fig7}(a) no spin bath included and in Fig.~\ref{fig7}(b) symmetric noise conditions were included with $b_{A}=b_{B}=0.5g$ and $g\tau_{A}=g\tau_{B}=0.5$. The results evidence that if we reduce the hyperfine coupling between the $e_{j}-\nu_{j}$ spins the numerical solutions go identical to the analytical results, validating the above results.

\begin{figure}[t]
\begin{center}
\includegraphics[width=0.48\textwidth]{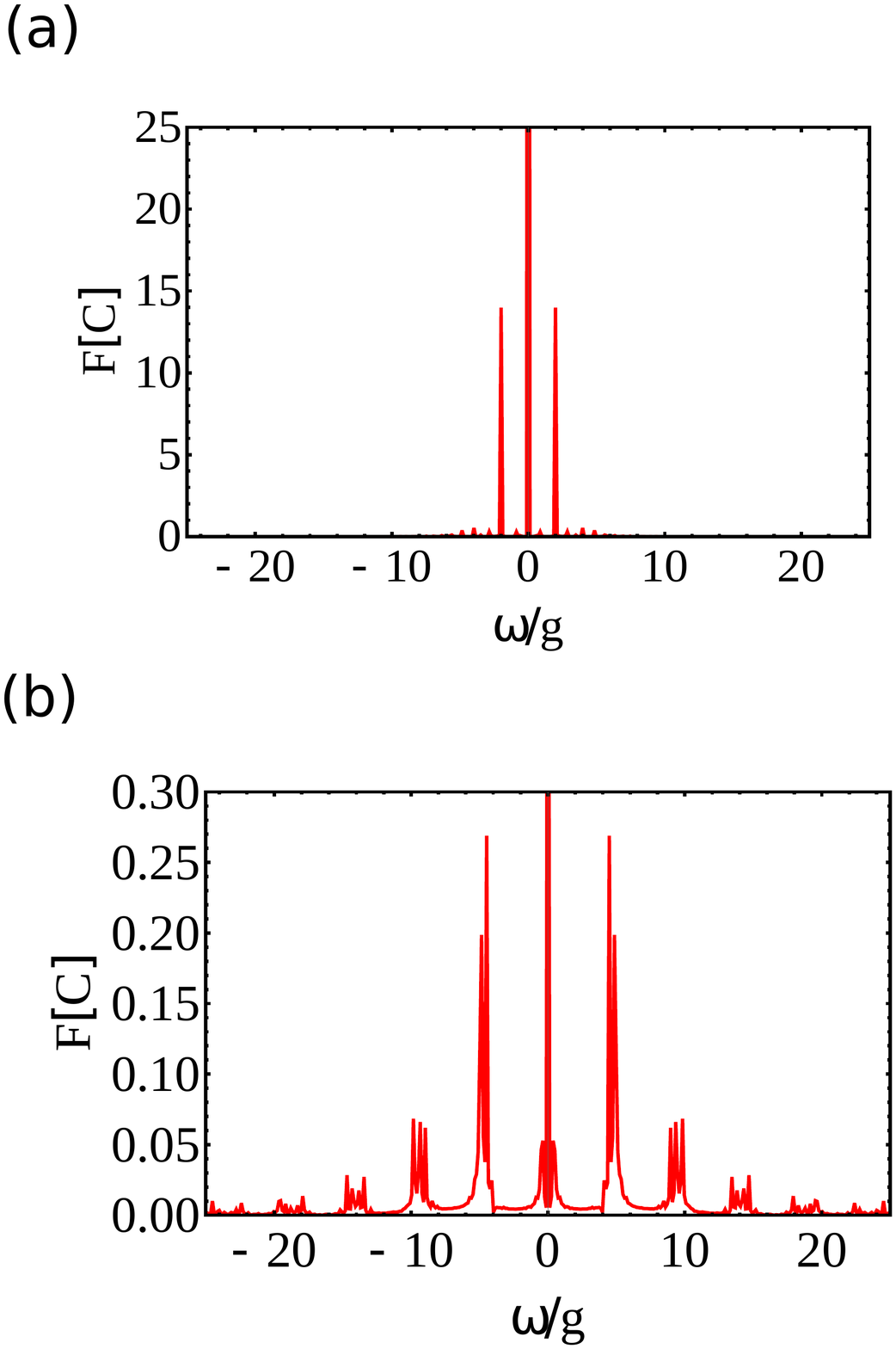}
\end{center}
\caption{Fourier transform of the concurrence $C$ as a function of dimensionless frequency $\omega/g$, symmetric case $g_{0,A}=g_{0,B}$, for $r=0.87$ and no spin bath was included. Panel (a) denote $e_{A}-e_{B}$, panel (b) represent $\nu_{A}-\nu_{B}$.
}
\label{fig8}
\end{figure}

\begin{figure}[t]
\begin{center}
\includegraphics[width=0.35\textwidth]{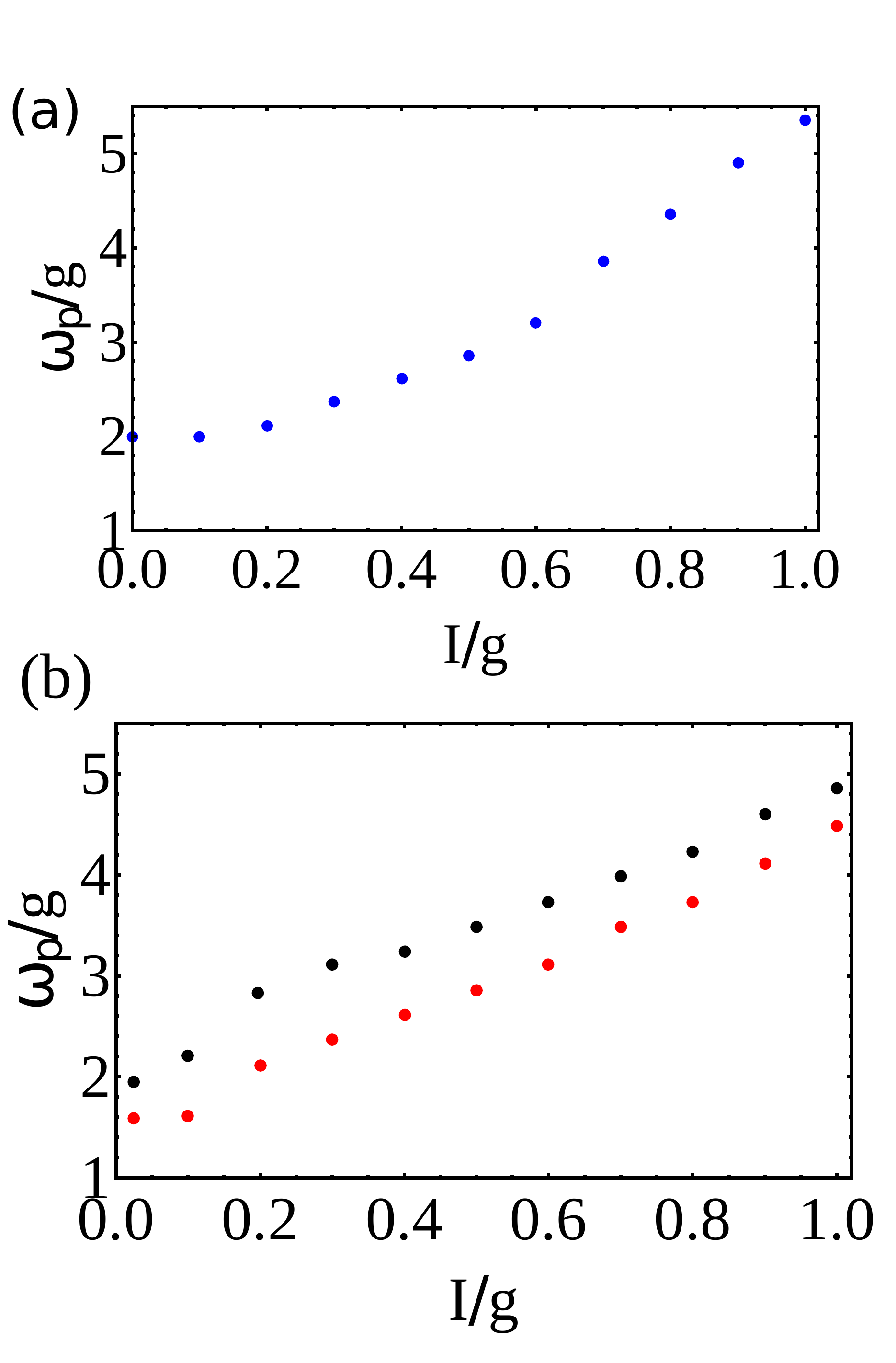}
\end{center}
\caption{Peak position of frequency in the Fourier transform of concurrence between spin pairs as a function of the hyperfine interaction $A(\vec{r})=B(\vec{r})=I$. (a) $e_{A}-e_{B}$ spins. (b) $\nu_{A}-\nu_{B}$ spins
}
\label{fig9}
\end{figure}

Now, we want highlight two elements of the presented results: first, we note that the nuclear entanglement persists for longer times compared with the electron entanglement even under noise environments. Second, in Figs.~\ref{fig2}-\ref{fig5} we observe very definite frequencies for the entanglement evolution in each system: $e_{j}-e_{j}$, $\nu_{j}-\nu_{j}$, $e_{j}-\nu_{j}$. Therefore, we calculate the Fourier transform of the concurrence in order to determine relevant frequencies in the system's entanglement dynamics. In Fig.~\ref{fig8} we show the results for the Fourier transform of the concurrence between $e_{A}-e_{B}$ (Fig.~\ref{fig8}(a)) and $\nu_{A}-\nu_{B}$(Fig.~\ref{fig8}(b)) as a function of the frequency $\omega$ in $g$ units, with symmetric conditions $g_{0,A}=g_{0,B}=g$ and no spin bath. The squeezing parameter $r$ was fixed as $r=0.87$ because we note that the central frequency in the Fourier transform does not change with the squeezing of the microwaves.
Besides that, in the nuclear entanglement we have a greater spectrum of relevant frequencies compared with the $e_{A}-e_{B}$ entanglement where the frequency appears as a more defined peak.
Finally, we present how to change the position of the peaks in the frequency scale ($\omega_{p}$) of the Fourier transform for $e_{A}-e_{B}$ (Fig.~\ref{fig9}(a)) and $\nu_{A}-\nu_{B}$ (Fig.~\ref{fig9}(b)) by varying the hyperfine coupling $A(\vec{r})=B(\vec{r})=I$. The results show a high dependence with the hyperfine coupling, and additional they recover the expected result for the uncoupled case $A(\vec{r})=B(\vec{r})=0$ where $\omega_{p}=2$

\section{Conclusions}
\label{sec:Conclusions}
In summary, we have derived analytical expressions for the density matrix describing the dynamics of distant electronic spins interacting  with a two mode squeezed state in a noise environment. We have characterized the dynamical entanglement in terms of the concurrence for the two spins approximating the effect of the bath, with a  classical theory, as a Ornstein-Uhlenbeck process. From our analytical and numerical results, we conclude that a squeezed microwave field produced by a parametric amplifier can be efficiently employed to induce entanglement in initially uncorrelated spin systems even when embedded in a noisy environment. We performed numerical simulations with the same initial states by varying the QMF and noise parameters, and obtained qualitatively
similar results.

The proposed scheme allows to evidence as the inclusion of noise environments change the optimal $r$ values to obtain maximum entanglement. In a realistic scenario, we have included the hyperfine interaction between the proximal $^{14}N$ spin and the electronic spin. In this situation the analytical expressions are not valid then a numerical solution was realized.

We extend our calculations to nuclear spins and electron-nucleus entanglement. Our result probes that even for nuclear spins which no interact directly with the entangled microwave field is possible an effective transfer of correlations mediated by the hyperfine electron-nuclear interaction. Besides for the nuclear systems the entanglement persist in spin baths environments that produce decoherence. While maximum entanglement is reached for small squeezing values for the electronic spins highly entangled states for the microwaves is required to entangle nuclear spins in a spin bath.

A shifting in the squeezing value for obtain maximum entanglement was shown for the electronic spins in presence of a spin bath, while for nuclear spins this value is constant. Moreover this scheme show the required values of squeezing in the studied systems and the limiting values for get entangled states in a spin bath.

Finally, we show that other quantum correlations besides entanglement persist even in noise environments and the effect of the spin bath is small on other correlations beyond entanglement.

\acknowledgements

\par A.V.G., F.J.R. and L.Q. acknowledge financial support from Facultad de Ciencias at UniAndes-2015 project "Transfer of correlations from non-classically correlated reservoirs to solid state systems" and project "Quantum control of non-equilibrium hybrid systems-Part II", UniAndes-2015.

\appendix
\section{Entanglement dynamics formalism for NV electronic spins}
\label{sec:appendixa}
It is well known that the Hamiltonian Eq.($\ref{Eqq:m8}$) commutes with the operator associated to the total number of excitations $\hat{{\cal N}}=\sum_{j=A,B}\left [ \hat{a_{j}}^{\dag }\hat{a_{j}}%
+\left( \frac{\hat{\sigma}_{z,j}+1}{2}\right)\right ].$ From this symmetry it follows
that the full spin-QMF Hilbert space can be separated in invariant
sub-spaces of dimension $2$ for each arm
\begin{equation}
\hat{H_{j}}=\sum_{n}\oplus \hat{H}_{n,j},
\label{eqa:2}
\end{equation}%
each sub-space spanned by orthonormal bases with $n_{j}$ excitations $\left\{
\left\vert \left( n_{j}-1\right) +\right\rangle ,\left\vert \left( n_{j}-1\right)
-\right\rangle \right\} $ expressed as:%
\begin{eqnarray}
\left\vert \left( n_{j}-1\right) +\right\rangle &=&\cos \left( \frac{\alpha _{n,j}%
}{2}\right) \left\vert n_{j}-1,e_{e}\right\rangle +\sin \left( \frac{\alpha _{n,j}}{2}%
\right) \left\vert n_{j},e_{g}\right\rangle,  \nonumber \\
\left\vert \left( n_{j}-1\right) -\right\rangle &=&-\cos \left( \frac{\alpha _{n,j}%
}{2}\right) \left\vert n_{j}-1,e_{e}\right\rangle +\sin \left( \frac{\alpha _{n,j}}{2}%
\right) \left\vert n_{j},e_{g}\right\rangle,  \nonumber\\
\label{eqa:3}
\end{eqnarray}%
with tan$\left( \alpha _{n,j}\right) =\frac{g_{0,j}\sqrt{n_{j}}}{\delta_j }$, the detuning is given by $\delta_j=\omega_j-\Omega_j$, and $\left\vert e_{g}\right\rangle$, $\left\vert e_{e}\right\rangle$, represent the ground and excited states for the electronic spin. This
latter symmetry can also be exploited by associating a su(2)-Lie algebra
within each invariant sub-space with $n$ total excitations as
\begin{equation}
\hat{J}_{x,j}=\frac{1}{2\sqrt{\hat{{\cal N}_{j}}}}\left( \hat{a}_{j}^{\dag }\sigma
_{j}^{-}+\hat{a}_{j}\sigma _{j}^{+}\right),
\label{eqa:4}
\end{equation}
\begin{equation}
\hat{J}_{y,j}=\frac{i}{2\sqrt{\hat{{\cal N}_{j}}}}\left( \hat{a}_{j}^{\dag }\sigma
_{j}^{-}-\hat{a}_{j}\sigma _{j}^{+}\right),
\label{eqa:5}
\end{equation}
\begin{equation}
\hat{J}_{z,j}=\frac{1}{2}\sigma _{z,j}.
\label{eqa:6}
\end{equation}
Therefore, the Hamiltonian for the sub-space with ${\cal N}_{j}$ excitations can be written as
\begin{equation}
\hat{H}(t)=\sum_{j=A,B}\Omega _{j}\hat{{\cal N}_{j}}+\delta _{j}\hat{J}_{z,j}+2%
\sqrt{n_{j}}g_{j}(t)\hat{J}_{x,j}-\frac{\Omega _{j}}{2}.
\label{eqa:6}
\end{equation}
In the interaction picture the Hamiltonian in
Eq.(\ref{eqa:6}) describes an effective spin in a time-dependent magnetic field
\begin{equation}
\hat{H_{j}}(t)=\sum_{j=A,B} \vec{\hat{J}}_{j}\cdot \vec{B}_{j}(t),
\label{eqa:7}
\end{equation}
with
\begin{equation}
\vec{B}_{j}(t)=\left( 2\sqrt{\hat{{\cal N}_{j}}}g_{j}(t),0,\delta _{j}\right).
\label{eqa:8}
\end{equation}
From now on we restrict to the resonance case $\delta_{j}=0$ yielding to a time-dependent field in the $x$-direction. Under this latter assumption the Hamiltonian commutes with itself at different times, leading to an exactly solvable evolution operator
\begin{equation}
\hat{U}_{j}(t)=e^{i2\theta _{j,n}(t)\hat{J}_{x,j}},
\label{eqa:9}
\end{equation}
with
\begin{equation}
\theta _{j,n}(t)=\sqrt{n_{j}}\int_{0}^{t}dt_{j}g(t_{j}).
\label{eqa:10}
\end{equation}
Note specially that at resonance
\begin{eqnarray}
\left\vert \left( n_{j}-1\right) ,+\right\rangle &=&\frac{1}{\sqrt{2}}\left[
\left\vert \left( n_{j}-1\right) ,e_{e}\right\rangle +\left\vert n_{j},e_{g}\right\rangle %
\right], \nonumber \\
\left\vert \left( n_{j}-1\right) ,-\right\rangle &=&\frac{1}{\sqrt{2}}\left[
-\left\vert \left( n_{j}-1\right) ,e_{e}\right\rangle +\left\vert n_{j},e_{g}\right\rangle %
\right]. \nonumber \\
\label{eqa:11}
\end{eqnarray}
Within the sub-space with ${\cal N}_{j}$ excitations it holds that
\begin{equation}
\hat{J}_{x,j}\left\vert \left( n_{j}-1\right) +\right\rangle =\frac{1}{2}%
\left\vert \left( n_{j}-1\right) +\right\rangle,
\label{eqa:12}
\end{equation}%
\begin{equation}
\hat{J}_{x,j}\left\vert \left( n_{j}-1\right) -\right\rangle =-\frac{1}{2}%
\left\vert \left( n_{j}-1\right) -\right\rangle.
\label{eqa:13}
\end{equation}
Eq.(\ref{eqa:9}) acting on the initial state for the system $\left\vert \psi(t=0) \right\rangle=\left\vert r\right\rangle\left\vert e_{g}\right\rangle_{A} \left\vert e_{g}\right\rangle_{B}$ (with $r$ given by the Eq.(\ref{eqa:14})), and properties in Eqs.(\ref{eqa:12})-(\ref{eqa:13}), allow us to easily obtain the NV-cavity quantum state at time $t$ as
\begin{equation}
\left\vert \psi (t)\right\rangle =\hat{U}_{A,B}(t)r_{c}\sum_{n=0}^{\infty }r_{t}^{n}\left\vert n,g\right\rangle _{A}\otimes \left\vert
n,g\right\rangle _{B},
\label{eqa:15}
\end{equation}
where we have written $r_{c}=1/$cosh(r) and $r_{t}=$tanh(r). In order to proceed further, individual terms in Eq.(\ref{eqa:15}) can be developed as
\begin{widetext}
\begin{equation}
\hat{U}_{j}(t)\left\vert n,g\right\rangle _{j}=r_{c}\sum_{n=0}^{\infty
}r_{t}^{n}\frac{1}{\sqrt{2}}\left[ e^{i2\sqrt{n_{j}}\theta _{j,n}(t)}\left\vert
\left( n_{j}-1\right) +\right\rangle +e^{-i2\sqrt{n_{j}}\theta _{j,n}}\left\vert
\left( n_{j}-1\right) -\right\rangle \right] _{j},
\label{eqa:16}
\end{equation}
\end{widetext}
with $\theta_{j,n}(t)$ given by Eq.(\ref{eqa:10}).
\section{Two-NV full density matrix}
\label{sec:appendix1}
Here we summarize some important intermediate steps to reach the analytical expression for the reduced two NV density matrix.
The density operator at time $t$ becomes
\begin{widetext}
\begin{eqnarray}
r_{c}^{2}\bar{\rho}(t) &=&\left\vert e_{g},0\right\rangle _{AA}\left\langle
e_{g},0\right\vert \otimes \left\vert e_{g},0\right\rangle _{BB}\left\langle
e_{g},0\right\vert + \nonumber \\
&&+\frac{1}{2}\sum_{n=1}^{\infty }r_{t}^{n}\left[ \left\langle e^{-i\theta
_{A,n}(t)}\right\rangle \left\vert e_{g},0\right\rangle _{AA}\left\langle \left(
n-1\right) +\right\vert +\left\langle e^{i\theta _{A,n}(t)}\right\rangle
\left\vert e_{g},0\right\rangle _{AA}\left\langle \left( n-1\right) -\right\vert %
\right] \otimes \nonumber \\
&&\otimes \left[ \left\langle e^{-i\theta _{B,n}(t)}\right\rangle \left\vert
e_{g},0\right\rangle _{BB}\left\langle \left( n-1\right) +\right\vert
+\left\langle e^{i\theta _{B,n}(t)}\right\rangle \left\vert e_{g},0\right\rangle
_{BB}\left\langle \left( n-1\right) -\right\vert \right] + \nonumber \\
&&+\frac{1}{2}\sum_{n=1}^{\infty }r_{t}^{n}\left[ \left\langle e^{i\theta
_{A,n}(t)}\right\rangle \left\vert \left( n-1\right) +\right\rangle
_{AA}\left\langle e_{g},0\right\vert +\left\langle e^{-i\theta
_{A,n}(t)}\right\rangle \left\vert \left( n-1\right) -\right\rangle
_{AA}\left\langle e_{g},0\right\vert \right] \otimes  \nonumber \\
&&\otimes \left[ \left\langle e^{i\theta _{B,n}(t)}\right\rangle \left\vert
\left( n-1\right) +\right\rangle _{BB}\left\langle e_{g},0\right\vert
+\left\langle e^{-i\theta _{B,n}(t)}\right\rangle \left\vert \left(
n-1\right) -\right\rangle _{BB}\left\langle e_{g},0\right\vert \right] + \nonumber \\
&&\frac{1}{4}\sum_{n=1}^{\infty }\sum_{m=1}^{\infty }r_{t}^{n+m}\left[
\left\langle e^{i\left( \theta _{A,n}(t)-\theta _{A,m}(t)\right)
}\right\rangle \left\vert \left( n-1\right) +\right\rangle _{AA}\left\langle
\left( m-1\right) ,+\right\vert \right. + \nonumber \\
&&+\left\langle e^{i\left( \theta _{A,n}(t)+\theta _{A,m}(t)\right)
}\right\rangle \left\vert \left( n-1\right) +\right\rangle _{AA}\left\langle
\left( m-1\right) ,-\right\vert + \nonumber \\
&&+\left\langle e^{-i\left( \theta _{A,n}(t)+\theta _{A,m}(t)\right)
}\right\rangle \left\vert \left( n-1\right) -\right\rangle _{AA}\left\langle
\left( m-1\right) ,+\right\vert + \nonumber \\
&&+\left. \left\langle e^{-i\left( \theta _{A,n}(t)-\theta _{A,m}(t)\right)
}\right\rangle \left\vert \left( n-1\right) -\right\rangle _{AA}\left\langle
\left( m-1\right) ,-\right\vert \right] \otimes \nonumber \\
&&\left[ \left\langle e^{i\left( \theta _{B,n}(t)-\theta _{B,m}(t)\right)
}\right\rangle \left\vert \left( n-1\right) +\right\rangle _{BB}\left\langle
\left( m-1\right) ,+\right\vert \right. + \nonumber \\
&&+\left\langle e^{i\left( \theta _{B,n}(t)+\theta _{B,m}(t)\right)
}\right\rangle \left\vert \left( n-1\right) +\right\rangle _{BB}\left\langle
\left( m-1\right) ,-\right\vert + \nonumber \\
&&+\left\langle e^{-i\left( \theta _{B,n}(t)+\theta _{B,m}(t)\right)
}\right\rangle \left\vert \left( n-1\right) -\right\rangle _{BB}\left\langle
\left( m-1\right) ,+\right\vert + \nonumber \\
&&+\left. \left\langle e^{-i\left( \theta _{B,n}(t)-\theta _{B,m}(t)\right)
}\right\rangle \left\vert \left( n-1\right) -\right\rangle _{BB}\left\langle
\left( m-1\right) ,-\right\vert \right].
\label{eqa:17}
\end{eqnarray}
\end{widetext}

The following expressions are valuable for that purpose:%

\begin{widetext}
\begin{eqnarray}
\sum_{p=0}^{\infty }\left\langle p\right\vert \left\vert \left( n-1\right)
+\right\rangle \left\langle \left( m-1\right) +\right\vert \left\vert
p\right\rangle  &=&\frac{1}{2}\left[ \delta _{n,m}\left( \left\vert
e_{g}\right\rangle \left\langle e_{g}\right\vert +\left\vert e_{e}\right\rangle
\left\langle e_{e}\right\vert \right) +\delta _{n-1,m}\left\vert e_{e}\right\rangle
\left\langle e_{g}\right\vert +\delta _{n,m-1}\left\vert e_{g}\right\rangle
\left\langle e_{e}\right\vert \right]  \nonumber \\
\sum_{p=0}^{\infty }\left\langle p\right\vert \left\vert \left( n-1\right)
+\right\rangle \left\langle \left( m-1\right) -\right\vert \left\vert
p\right\rangle  &=&\frac{1}{2}\left[ \delta _{n,m}\left( \left\vert
e_{g}\right\rangle \left\langle e_{g}\right\vert -\left\vert e_{e}\right\rangle
\left\langle e_{e}\right\vert \right) +\delta _{n-1,m}\left\vert e_{e}\right\rangle
\left\langle e_{g}\right\vert -\delta _{n,m-1}\left\vert e_{g}\right\rangle
\left\langle e_{e}\right\vert \right]   \nonumber \\
\sum_{p=0}^{\infty }\left\langle p\right\vert \left\vert \left( n-1\right)
-\right\rangle \left\langle \left( m-1\right) +\right\vert \left\vert
p\right\rangle  &=&\frac{1}{2}\left[ \delta _{n,m}\left( \left\vert
e_{g}\right\rangle \left\langle e_{g}\right\vert -\left\vert e_{e}\right\rangle
\left\langle e_{e}\right\vert \right) -\delta _{n-1,m}\left\vert e_{e}\right\rangle
\left\langle e_{g}\right\vert +\delta _{n,m-1}\left\vert e_{g}\right\rangle
\left\langle e_{e}\right\vert \right] \nonumber  \\
\sum_{p=0}^{\infty }\left\langle p\right\vert \left\vert \left( n-1\right)
-\right\rangle \left\langle \left( m-1\right) -\right\vert \left\vert
p\right\rangle  &=&\frac{1}{2}\left[ \delta _{n,m}\left( \left\vert
e_{g}\right\rangle \left\langle e_{g}\right\vert +\left\vert e_{e}\right\rangle
\left\langle e_{e}\right\vert \right) -\delta _{n-1,m}\left\vert e_{e}\right\rangle
\left\langle e_{g}\right\vert +\delta _{n,m-1}\left\vert e_{g}\right\rangle
\left\langle e_{e}\right\vert \right].
\label{eqa:18}
\end{eqnarray}
\end{widetext}

\section{Density matrix for electronic NV spins with constant spin-cavity couplings}
\label{sec:appendixB}
In order to calculate the density matrix elements in the case where the spin-cavity coupling is constant, we have evaluated the expression Eq.(\ref{ecrho2}) with $g_{0,A}=g_{0,B}=g$,  in this limit we have
\begin{equation}
\rho _{1,1}(t)=\frac{1}{r_{c}^{2}}\sum_{n=0}^{\infty }r_{t}^{2n}\cos
^{4}\left( \sqrt{n}g_{0}t\right),
\label{Eqr1}
\end{equation}

\begin{equation}
\rho _{2,2}(t)=\rho _{3,3}(t)=\frac{1}{r_{c}^{2}}\sum_{n=0}^{\infty
}r_{t}^{2n}\sin ^{2}\left( \sqrt{n}g_{0}t\right) \cos ^{2}\left( \sqrt{n}%
g_{0}t\right),
\label{Eqr2}
\end{equation}

\begin{equation}
\rho _{4,4}(t)=\frac{1}{r_{c}^{2}}\sum_{n=0}^{\infty }r_{t}^{2n}\sin
^{4}\left( \sqrt{n}g_{0}t\right),
\label{Eqr3}
\end{equation}
and the non-diagonal term becomes%
\begin{equation}
\rho _{1,4}(t)=-\frac{1}{r_{c}^{2}}\sum_{n=0}^{\infty }r_{t}^{2n+1}\sin
^{2}\left( \sqrt{n+1}g_{0}t\right) \cos ^{2}\left( \sqrt{n}g_{0}t\right).
\label{Eqr4}
\end{equation}

\section{Time dependent coefficients}
\label{sec:appendix2}
In this section we provide the elements of the reduced $16\times 16$ density matrix for the electronic and nuclear spins in a noisy environment. In the main text we present simplified analytical expression for two electronic spins when no hyperfine coupling with the proximal nuclear spin. However a numerical solution is needed if we include this interaction and the spin $^{13}C$ bath.
We start defining the systems's state at time $t$ as
\begin{widetext}
\begin{eqnarray}
\left\vert \psi (t)\right\rangle &=&\hat{U}_{A,B}(t)\sum_{n=0}^{\infty
}\alpha _{n}\left\vert n,e_{g},\nu _{g}\right\rangle _{A}\otimes \left\vert
n,e_{g},\nu _{g}\right\rangle _{B}  \label{Eq58} \\
&=&\sum_{n=0}^{\infty }\alpha _{n}\left[ \hat{U}_{A}(t)\left\vert
n,e_{g},\nu _{g}\right\rangle _{A}\right] \otimes \left[ \hat{U}%
_{B}(t)\left\vert n,e_{g},\nu _{g}\right\rangle _{B}\right]  \nonumber \\
&=&\alpha _{0}\left\vert n,e_{g},\nu _{g}\right\rangle _{A}\otimes
\left\vert n,e_{g},\nu _{g}\right\rangle _{B}  \nonumber \\
&&+\alpha _{1}\left[ \sum_{i=1}^{3}C_{1,i}(t)\left\vert 1,i\right\rangle %
\right] _{A}\otimes \left[ \sum_{i=1}^{3}C_{1,j}(t)\left\vert
1,j\right\rangle \right] _{B}+\sum_{N=2}^{\infty }\alpha _{N}\left[
\sum_{i=1}^{4}C_{N,i}(t)\left\vert N,i\right\rangle \right] _{A}\otimes %
\left[ \sum_{j=1}^{4}C_{N,j}(t)\left\vert N,j\right\rangle \right] _{B}
\nonumber
\end{eqnarray}%
\end{widetext}
where
\begin{equation}
\left\vert \alpha _{N}\right\vert=\frac{tanh(r)^{N}}{cosh(r)}.
\end{equation}
The terms $C_{1,i}(t)$, $C_{1,j}(t)$ (where $i$ and $j$ can take values $1,2,3$) are the coefficients at time $t$ in the expansion for the state in the sub-space with $N=1$ excitations in the branches $A$ and $B$, respectively. The coefficients $C_{N,i}(t)$ and $C_{N,j}(t)$ (where $i$ and $j$ in this case can take values $1,2,3,4$) allow determine the state at time $t$ in the four dimensional subspaces with $N\geq 2$. Now, we can proceed to evaluate the density matrix as $\hat{\rho}%
(t)=\left\vert \psi (t)\right\rangle \left\langle \psi (t)\right\vert$ and the reduced density matrix tracing over the state of the field
\begin{eqnarray}
\bar{\rho}_{2Q}(t)=\sum_{p=0}^{\infty }\sum_{q=0}^{\infty }\left\langle
p,q\right\vert \bar{\rho}(t)\left\vert p,q\right\rangle,
\label{a11}
\end{eqnarray}
with $p$ and $q$ the photon number in the two branches. The  bar in $\bar{\rho}_{2Q}(t)$ and $\bar{\rho}(t)$ represent stochastic terms due to the noise spin bath. The diagonal elements obtained for the density matrix are given by
\begin{widetext}
\begin{eqnarray}
\rho _{1,1}&=&\left\vert \alpha _{0}\right\vert ^{2}+\left\vert \alpha
_{1}\right\vert ^{2}\left\langle \left\vert C_{1,2}(t)\right\vert
^{2}\right\rangle _{A}\left\langle \left\vert C_{1,2}(t)\right\vert
^{2}\right\rangle _{B}+\sum_{N=2}^{\infty }\left\vert \alpha _{N}\right\vert
^{2}\left\langle \left\vert C_{N,2}(t)\right\vert ^{2}\right\rangle
_{A}\left\langle \left\vert C_{N,2}(t)\right\vert ^{2}\right\rangle _{B} \notag \\
\end{eqnarray}
\end{widetext}

\begin{widetext}
\begin{eqnarray}
\rho _{2,2}&=&\left\vert \alpha _{1}\right\vert ^{2}\left\langle \left\vert
C_{1,2}(t)\right\vert ^{2}\right\rangle _{A}\left\langle \left\vert
C_{1,3}(t)\right\vert ^{2}\right\rangle _{B}+\sum_{N=2}^{\infty }\left\vert
\alpha _{N}\right\vert ^{2}\left\langle \left\vert C_{N,2}(t)\right\vert
^{2}\right\rangle _{A}\left\langle \left\vert C_{N,4}(t)\right\vert
^{2}\right\rangle _{B} \notag \\
\end{eqnarray}
\end{widetext}

\begin{widetext}
\begin{eqnarray}
\rho _{3,3}&=&\left\vert \alpha _{1}\right\vert ^{2}\left\langle \left\vert
C_{1,2}(t)\right\vert ^{2}\right\rangle _{A}\left\langle \left\vert
C_{1,1}(t)\right\vert ^{2}\right\rangle _{B}+\sum_{N=2}^{\infty }\left\vert
\alpha _{N}\right\vert ^{2}\left\langle \left\vert C_{N,2}(t)\right\vert
^{2}\right\rangle _{A}\left\langle \left\vert C_{N,1}(t)\right\vert
^{2}\right\rangle _{B} \notag \\
\end{eqnarray}
\end{widetext}

\begin{widetext}
\begin{eqnarray}
\rho _{4,4}&=&\sum_{N=2}^{\infty }\left\vert \alpha _{N}\right\vert
^{2}\left\langle \left\vert C_{N,2}(t)\right\vert ^{2}\right\rangle
_{A}\left\langle \left\vert C_{N,3}(t)\right\vert ^{2}\right\rangle _{B}  \notag \\
\rho _{5,5}&=&\left\vert \alpha _{1}\right\vert ^{2}\left\langle \left\vert
C_{1,3}(t)\right\vert ^{2}\right\rangle _{A}\left\langle \left\vert
C_{1,2}(t)\right\vert ^{2}\right\rangle _{B}+\sum_{N=2}^{\infty }\left\vert
\alpha _{N}\right\vert ^{2}\left\langle \left\vert C_{N,4}(t)\right\vert
^{2}\right\rangle _{A}\left\langle \left\vert C_{N,2}(t)\right\vert
^{2}\right\rangle _{B} \notag \\
\rho _{6,6}&=&\left\vert \alpha _{1}\right\vert ^{2}\left\langle \left\vert
C_{1,3}(t)\right\vert ^{2}\right\rangle _{A}\left\langle \left\vert
C_{1,3}(t)\right\vert ^{2}\right\rangle _{B}+\sum_{N=2}^{\infty }\left\vert
\alpha _{N}\right\vert ^{2}\left\langle \left\vert C_{N,4}(t)\right\vert
^{2}\right\rangle _{A}\left\langle \left\vert C_{N,4}(t)\right\vert
^{2}\right\rangle _{B} \notag \\
\rho _{7,7}&=&\left\vert \alpha _{1}\right\vert ^{2}\left\langle \left\vert
C_{1,3}(t)\right\vert ^{2}\right\rangle _{A}\left\langle \left\vert
C_{1,1}(t)\right\vert ^{2}\right\rangle _{B}+\sum_{N=2}^{\infty }\left\vert
\alpha _{N}\right\vert ^{2}\left\langle \left\vert C_{N,4}(t)\right\vert
^{2}\right\rangle _{A}\left\langle \left\vert C_{N,1}(t)\right\vert
^{2}\right\rangle _{B} \notag \\
\rho _{8,8}&=&\sum_{N=2}^{\infty }\left\vert \alpha _{N}\right\vert
^{2}\left\langle \left\vert C_{N,4}(t)\right\vert ^{2}\right\rangle
_{A}\left\langle \left\vert C_{N,3}(t)\right\vert ^{2}\right\rangle _{B} \notag \\
\rho _{9,9}&=&\left\vert \alpha _{1}\right\vert ^{2}\left\langle \left\vert
C_{1,1}(t)\right\vert ^{2}\right\rangle _{A}\left\langle \left\vert
C_{1,2}(t)\right\vert ^{2}\right\rangle _{B}+\sum_{N=2}^{\infty }\left\vert
\alpha _{N}\right\vert ^{2}\left\langle \left\vert C_{N,1}(t)\right\vert
^{2}\right\rangle _{A}\left\langle \left\vert C_{N,2}(t)\right\vert
^{2}\right\rangle _{B} \notag \\
\rho _{10,10}&=&\left\vert \alpha _{1}\right\vert ^{2}\left\langle \left\vert
C_{1,1}(t)\right\vert ^{2}\right\rangle _{A}\left\langle \left\vert
C_{1,3}(t)\right\vert ^{2}\right\rangle _{B}+\sum_{N=2}^{\infty }\left\vert
\alpha _{N}\right\vert ^{2}\left\langle \left\vert C_{N,1}(t)\right\vert
^{2}\right\rangle _{A}\left\langle \left\vert C_{N,4}(t)\right\vert
^{2}\right\rangle _{B} \notag \\
\rho _{11,11}&=&\left\vert \alpha _{1}\right\vert ^{2}\left\langle \left\vert
C_{1,1}(t)\right\vert ^{2}\right\rangle _{A}\left\langle \left\vert
C_{1,1}(t)\right\vert ^{2}\right\rangle _{B}+\sum_{N=2}^{\infty }\left\vert
\alpha _{N}\right\vert ^{2}\left\langle \left\vert C_{N,1}(t)\right\vert
^{2}\right\rangle _{A}\left\langle \left\vert C_{N,1}(t)\right\vert
^{2}\right\rangle _{B} \notag \\
\rho _{12,12}&=&\sum_{N=2}^{\infty }\left\vert \alpha _{N}\right\vert
^{2}\left\langle \left\vert C_{N,1}(t)\right\vert ^{2}\right\rangle
_{A}\left\langle \left\vert C_{N,3}(t)\right\vert ^{2}\right\rangle _{B} \notag \\
\rho _{13,13}&=&\sum_{N=2}^{\infty }\left\vert \alpha _{N}\right\vert
^{2}\left\langle \left\vert C_{N,3}(t)\right\vert ^{2}\right\rangle
_{A}\left\langle \left\vert C_{N,2}(t)\right\vert ^{2}\right\rangle _{B} \notag \\
\rho _{14,14}&=&\sum_{N=2}^{\infty }\left\vert \alpha _{N}\right\vert
^{2}\left\langle \left\vert C_{N,3}(t)\right\vert ^{2}\right\rangle
_{A}\left\langle \left\vert C_{N,4}(t)\right\vert ^{2}\right\rangle _{B} \notag \\
\rho _{15,15}&=&\sum_{N=2}^{\infty }\left\vert \alpha _{N}\right\vert
^{2}\left\langle \left\vert C_{N,3}(t)\right\vert ^{2}\right\rangle
_{A}\left\langle \left\vert C_{N,1}(t)\right\vert ^{2}\right\rangle _{B} \notag \\
\rho _{16,16}&=&\sum_{N=2}^{\infty }\left\vert \alpha _{N}\right\vert
^{2}\left\langle \left\vert C_{N,3}(t)\right\vert ^{2}\right\rangle
_{A}\left\langle \left\vert C_{N,3}(t)\right\vert ^{2}\right\rangle _{B}
\label{Eq60}
\end{eqnarray}
\end{widetext}
where $N$ vary between $0$ and the photon number state in the field. In our simulation we have considered $N=85$ and we have probed the essential conditions for a density matrix. The results yield to $Tr\left\{ \bar{\rho}_{2Q}(t)\right\} =1$ as it should be. The stochastic realizations in the coefficients $\left\langle  ...\right\rangle$ take into account many realizations in the systems when we include the noise parameters. In our calculation we have evaluated the average taking approximately $10000$ realizations in the coefficients average. Symmetric conditions have been considered in the two branches. Non-diagonal elements in the density matrix are:
\begin{widetext}
\begin{eqnarray}
\rho _{1,6}&=&\rho _{6,1}^{\ast }=\alpha _{0}\alpha _{1}^{\ast }\left\langle
C_{1,3}(t)\right\rangle _{A}\left\langle C_{1,3}(t)^{\ast }\right\rangle
_{B}+\sum_{N=1}^{\infty }\alpha _{N}\alpha _{N+1}^{\ast }\left\langle
C_{N,2}(t)C_{N+1,4}^{\ast }(t)\right\rangle _{A}\left\langle
C_{N,2}(t)C_{N+1,4}^{\ast }(t)\right\rangle _{B} \notag \\
\rho _{1,7}&=&\rho _{7,1}^{\ast }=\alpha _{0}\alpha _{1}^{\ast }\left\langle
C_{1,3}(t)\right\rangle _{A}\left\langle C_{1,1}(t)^{\ast }\right\rangle
_{B}+\sum_{N=1}^{\infty }\alpha _{N}\alpha _{N+1}^{\ast }\left\langle
C_{N,2}(t)C_{N+1,4}^{\ast }(t)\right\rangle _{A}\left\langle
C_{N,2}(t)C_{N+1,1}^{\ast }(t)\right\rangle _{B} \notag \\
\rho _{1,10}&=&\rho _{10,1}^{\ast }=\alpha _{0}\alpha _{1}^{\ast }\left\langle
C_{1,1}(t)\right\rangle _{A}\left\langle C_{1,3}(t)^{\ast }\right\rangle
_{B}+\sum_{N=1}^{\infty }\alpha _{N}\alpha _{N+1}^{\ast }\left\langle
C_{N,2}(t)C_{N+1,1}^{\ast }(t)\right\rangle _{A}\left\langle
C_{N,2}(t)C_{N+1,4}^{\ast }(t)\right\rangle _{B} \notag \\
\rho _{1,11}&=&\rho _{11,1}^{\ast }=\alpha _{0}\alpha _{1}^{\ast }\left\langle
C_{1,1}(t)\right\rangle _{A}\left\langle C_{1,1}(t)^{\ast }\right\rangle
_{B}+\sum_{N=1}^{\infty }\alpha _{N}\alpha _{N+1}^{\ast }\left\langle
C_{N,2}(t)C_{N+1,1}^{\ast }(t)\right\rangle _{A}\left\langle
C_{N,2}(t)C_{N+1,1}^{\ast }(t)\right\rangle _{B} \notag \\
\rho _{1,16}&=&\rho _{16,1}^{\ast }=\sum_{N=0}^{\infty }\alpha _{N}\alpha
_{N+2}^{\ast }\left\langle C_{N,2}(t)C_{N+2,3}^{\ast }(t)\right\rangle
_{A}\left\langle C_{N,2}(t)C_{N+2,3}^{\ast }(t)\right\rangle _{B} \notag \\
\rho _{2,3}&=&\rho _{3,2}^{\ast }=\left\vert \alpha _{1}\right\vert
^{2}\left\langle \left\vert C_{1,2}(t)\right\vert ^{2}\right\rangle
_{A}\left\langle C_{1,3}(t)C_{1,1}(t)^{\ast }\right\rangle
_{B}+\sum_{N=2}^{\infty }\left\vert \alpha _{N}\right\vert ^{2}\left\langle
\left\vert C_{N,2}(t)\right\vert ^{2}\right\rangle _{A}\left\langle
C_{N,4}(t)C_{N,1}^{\ast }(t)\right\rangle _{B} \notag \\
\rho _{2,8}&=&\rho _{8,2}^{\ast }=\alpha _{1}\alpha _{2}^{\ast }\left\langle
C_{1,2}(t)C_{2,4}(t)^{\ast }\right\rangle _{A}\left\langle
C_{1,3}(t)C_{2,3}(t)^{\ast }\right\rangle _{B}+\sum_{N=2}^{\infty }\alpha
_{N}\alpha _{N+1}^{\ast }\left\langle C_{N,2}(t)C_{N+1,4}^{\ast
}(t)\right\rangle _{A}\left\langle C_{N,4}(t)C_{N+1,3}^{\ast
}(t)\right\rangle _{B}  \notag \\
\rho _{2,12}&=&\rho _{12,2}^{\ast }=\alpha _{1}\alpha _{2}^{\ast }\left\langle
C_{1,2}(t)C_{2,1}(t)^{\ast }\right\rangle _{A}\left\langle
C_{1,3}(t)C_{2,3}(t)^{\ast }\right\rangle _{B}+\sum_{N=2}^{\infty }\alpha
_{N}\alpha _{N+1}^{\ast }\left\langle C_{N,2}(t)C_{N+1,1}^{\ast
}(t)\right\rangle _{A}\left\langle C_{N,4}(t)C_{N+1,3}^{\ast
}(t)\right\rangle _{B}  \notag \\
\rho _{3,8}&=&\rho _{8,3}^{\ast }=\sum_{N=1}^{\infty }\alpha _{N}\alpha
_{N+1}^{\ast }\left\langle C_{N,2}(t)C_{N+1,4}^{\ast }(t)\right\rangle
_{A}\left\langle C_{N,1}(t)C_{N+1,3}^{\ast }(t)\right\rangle _{B} \notag \\
\rho _{3,12}&=&\rho _{12,3}^{\ast }=\sum_{N=1}^{\infty }\alpha _{N}\alpha
_{N+1}^{\ast }\left\langle C_{N,2}(t)C_{N+1,1}^{\ast }(t)\right\rangle
_{A}\left\langle C_{N,1}(t)C_{N+1,3}^{\ast }(t)\right\rangle _{B} \notag \\
\rho _{5,9}&=&\rho _{9,5}^{\ast }=\left\vert \alpha _{1}\right\vert
^{2}\left\langle C_{1,3}(t)C_{1,1}(t)^{\ast }\right\rangle _{A}\left\langle
\left\vert C_{1,2}(t)\right\vert ^{2}\right\rangle _{B}+\sum_{N=2}^{\infty
}\left\vert \alpha _{N}\right\vert ^{2}\left\langle C_{N,4}(t)C_{N,1}^{\ast
}(t)\right\rangle _{A}\left\langle \left\vert C_{N,2}(t)\right\vert
^{2}\right\rangle _{B}  \notag \\
\rho _{5,14}&=&\rho _{14,5}^{\ast }=\alpha _{1}\alpha _{2}^{\ast }\left\langle
C_{1,3}(t)C_{2,3}(t)^{\ast }\right\rangle _{A}\left\langle
C_{1,2}(t)C_{2,4}(t)^{\ast }\right\rangle _{B}+\sum_{N=2}^{\infty }\alpha
_{N}\alpha _{N+1}^{\ast }\left\langle C_{N,4}(t)C_{N+1,3}^{\ast
}(t)\right\rangle _{A}\left\langle C_{N,2}(t)C_{N+1,4}^{\ast
}(t)\right\rangle _{B}  \notag \\
\rho _{5,15}&=&\rho _{15,5}^{\ast }=\alpha _{1}\alpha _{2}^{\ast }\left\langle
C_{1,3}(t)C_{2,3}(t)^{\ast }\right\rangle _{A}\left\langle
C_{1,2}(t)C_{2,1}(t)^{\ast }\right\rangle _{B}+\sum_{N=2}^{\infty }\alpha
_{N}\alpha _{N+1}^{\ast }\left\langle C_{N,4}(t)C_{N+1,3}^{\ast
}(t)\right\rangle _{A}\left\langle C_{N,2}(t)C_{N+1,1}^{\ast
}(t)\right\rangle _{B} \notag \\
\rho _{6,7}&=&\rho _{7,6}^{\ast }=\left\vert \alpha _{1}\right\vert
^{2}\left\langle \left\vert C_{1,3}(t)\right\vert ^{2}\right\rangle
_{A}\left\langle C_{1,3}(t)C_{1,1}(t)^{\ast }\right\rangle
_{B}+\sum_{N=2}^{\infty }\left\vert \alpha _{N}\right\vert ^{2}\left\langle
\left\vert C_{N,4}(t)\right\vert ^{2}\right\rangle _{A}\left\langle
C_{N,4}(t)C_{N,1}^{\ast }(t)\right\rangle _{B} \notag \\
\rho _{6,10}&=&\rho _{10,6}^{\ast }=\sum_{N=2}^{\infty }\left\vert \alpha
_{N}\right\vert ^{2}\left\langle C_{N,4}(t)C_{N,1}^{\ast }(t)\right\rangle
_{A}\left\langle \left\vert C_{N,4}(t)\right\vert ^{2}\right\rangle _{B} \notag \\
\rho _{6,11}&=&\rho _{11,6}^{\ast }=\sum_{N=2}^{\infty }\left\vert \alpha
_{N}\right\vert ^{2}\left\langle C_{N,4}(t)C_{N,1}^{\ast }(t)\right\rangle
_{A}\left\langle C_{N,4}(t)C_{N,1}^{\ast }(t)\right\rangle _{B} \notag \\
\rho _{6,16}&=&\rho _{16,6}^{\ast }=\alpha _{1}\alpha _{2}^{\ast }\left\langle
C_{1,3}(t)C_{2,3}^{\ast }(t)\right\rangle _{A}\left\langle
C_{1,3}(t)C_{2,3}^{\ast }(t)\right\rangle _{B}+\sum_{N=2}^{\infty }\alpha
_{N}\alpha _{N+1}^{\ast }\left\langle C_{N,4}(t)C_{N+1,3}^{\ast
}(t)\right\rangle _{A}\left\langle C_{N,4}(t)C_{N+1,3}^{\ast
}(t)\right\rangle _{B}  \notag \\
\rho _{7,10}&=&\rho _{10,7}^{\ast }=\left\vert \alpha _{1}\right\vert
^{2}\left\langle C_{1,3}(t)C_{1,1}^{\ast }(t)\right\rangle _{A}\left\langle
C_{1,1}(t)C_{1,3}^{\ast }(t)\right\rangle _{B}+\sum_{N=2}^{\infty
}\left\vert \alpha _{N}\right\vert ^{2}\left\langle C_{N,4}(t)C_{N,1}^{\ast
}(t)\right\rangle _{A}\left\langle C_{N,1}(t)C_{N,4}^{\ast }(t)\right\rangle
_{B}  \notag \\
\rho _{7,11}&=&\rho _{11,7}^{\ast }=\left\vert \alpha _{1}\right\vert
^{2}\left\langle C_{1,3}(t)C_{1,1}^{\ast }(t)\right\rangle _{A}\left\langle
\left\vert C_{1,1}(t)\right\vert ^{2}\right\rangle _{B}+\sum_{N=2}^{\infty
}\left\vert \alpha _{N}\right\vert ^{2}\left\langle C_{N,4}(t)C_{N,1}^{\ast
}(t)\right\rangle _{A}\left\langle \left\vert C_{N,1}(t)\right\vert
^{2}\right\rangle _{B}  \notag \\
\rho _{7,16}&=&\rho _{16,7}^{\ast }=\alpha _{1}\alpha _{2}^{\ast }\left\langle
C_{1,3}(t)C_{2,3}^{\ast }(t)\right\rangle _{A}\left\langle
C_{1,1}(t)C_{2,3}^{\ast }(t)\right\rangle _{B}+\sum_{N=2}^{\infty }\alpha
_{N}\alpha _{N+1}^{\ast }\left\langle C_{N,4}(t)C_{N+1,3}^{\ast
}(t)\right\rangle _{A}\left\langle C_{N,1}(t)C_{N+1,3}^{\ast
}(t)\right\rangle _{B}  \notag \\
\rho _{8,12}&=&\rho _{12,8}^{\ast }=\sum_{N=2}^{\infty }\left\vert \alpha
_{N}\right\vert ^{2}\left\langle C_{N,4}(t)C_{N,1}^{\ast }(t)\right\rangle
_{A}\left\langle \left\vert C_{N,3}(t)\right\vert ^{2}\right\rangle _{B} \notag \\
\rho _{9,14}&=&\rho _{14,9}^{\ast }=\sum_{N=1}^{\infty }\alpha _{N}\alpha
_{N+1}^{\ast }\left\langle C_{N,1}(t)C_{N+1,3}^{\ast }(t)\right\rangle
_{A}\left\langle C_{N,2}(t)C_{N+1,4}^{\ast }(t)\right\rangle _{B} \notag \\
\rho _{9,15}&=&\rho _{15,9}^{\ast }=\sum_{N=1}^{\infty }\alpha _{N}\alpha
_{N+1}^{\ast }\left\langle C_{N,1}(t)C_{N+1,3}^{\ast }(t)\right\rangle
_{A}\left\langle C_{N,2}(t)C_{N+1,1}^{\ast }(t)\right\rangle _{B} \notag \\
\rho _{10,11}&=&\rho _{11,10}^{\ast }=\left\vert \alpha _{1}\right\vert
^{2}\left\langle C_{1,3}(t)C_{1,1}^{\ast }(t)\right\rangle _{A}\left\langle
C_{1,1}(t)C_{1,3}^{\ast }(t)\right\rangle _{B}+\sum_{N=2}^{\infty
}\left\vert \alpha _{N}\right\vert ^{2}\left\langle C_{N,4}(t)C_{N,1}^{\ast
}(t)\right\rangle _{A}\left\langle C_{N,1}(t)C_{N,4}^{\ast }(t)\right\rangle
_{B}  \notag \\
\rho _{10,16}&=&\rho _{16,10}^{\ast }=\alpha _{1}\alpha _{2}^{\ast
}\left\langle C_{1,1}(t)C_{2,3}^{\ast }(t)\right\rangle _{A}\left\langle
C_{1,3}(t)C_{2,3}^{\ast }(t)\right\rangle _{B}+\sum_{N=2}^{\infty }\alpha
_{N}\alpha _{N+1}^{\ast }\left\langle C_{N,1}(t)C_{N+1,3}^{\ast
}(t)\right\rangle _{A}\left\langle C_{N,4}(t)C_{N+1,3}^{\ast
}(t)\right\rangle _{B}  \notag \\
\rho _{11,16}&=&\rho _{16,11}^{\ast }=\sum_{N=1}^{\infty }\alpha _{N}\alpha
_{N+1}^{\ast }\left\langle C_{N,1}(t)C_{N+1,3}^{\ast }(t)\right\rangle
_{A}\left\langle C_{N,1}(t)C_{N+1,3}^{\ast }(t)\right\rangle _{B} \notag \\
\rho _{14,15}&=&\rho _{15,14}^{\ast }=\sum_{N=2}^{\infty }\left\vert \alpha
_{N}\right\vert ^{2}\left\langle \left\vert C_{N,3}(t)\right\vert
^{2}\right\rangle _{A}\left\langle C_{N,4}(t)C_{N,1}^{\ast }(t)\right\rangle
_{B}
\label{Eq61}
\end{eqnarray}
\end{widetext}

\bibliography{mybib}	
\end{document}